\documentclass[12pt, preprint]{aastex}
\usepackage{amsmath}
\usepackage{graphicx}
\usepackage{natbib}

\usepackage[colorlinks=true,citecolor=blue,breaklinks=true,linktocpage=true]{hyperref}
\bibpunct{(}{)}{;}{a}{}{,} 
\usepackage[pagewise]{lineno}
\usepackage{xspace}
\usepackage{hyperref}

\usepackage{multirow}

\newtheorem{conj}{Conjecture}

\def\vec#1{\ensuremath{\mathbf{#1}}}

\newcommand{\kc}{\ensuremath{k_{\rm c}}}
\newcommand{\Pc}{\ensuremath{P_{\rm c}}}
\newcommand{\kcl}{\ensuremath{k_{{\rm c}, l}}}
\newcommand{\ocl}{\ensuremath{\omega_{{\rm c}, l}}}
\newcommand{\nbarcl}{\ensuremath{\bar{n}_{{\rm c}, l}}}
\newcommand{\Dbarcl}{\ensuremath{\bar{D}_{{\rm c}, l}}}
\newcommand{\vph}{\ensuremath{v_{\rm ph}}}
\newcommand{\vgr}{\ensuremath{v_{\rm gr}}}
\newcommand{\rhoi}{\ensuremath{\rho_{\rm i}}}
\newcommand{\rhoe}{\ensuremath{\rho_{\rm e}}}
\newcommand{\vai}{\ensuremath{v_{\rm Ai}}}
\newcommand{\vae}{\ensuremath{v_{\rm Ae}}}
\newcommand{\vgmin}{\ensuremath{v_{\rm gr}^{\rm min}}}

\newcommand{\mathd}{\ensuremath{{\rm d}}}
\newcommand{\Ai}{\ensuremath{{\rm Ai}}}
\newcommand{\Bi}{\ensuremath{{\rm Bi}}}
\newcommand{\sech}{\ensuremath{{\rm sech}}}

\usepackage{color}

\shorttitle{Impulsively Generated Sausage Waves in Nonuniform Slabs}
\shortauthors{Li et al.}

\begin{document}


\title{IMPULSIVELY GENERATED WAVE TRAINS IN CORONAL STRUCTURES:
II. EFFECTS OF TRANSVERSE STRUCTURING ON SAUSAGE WAVES IN PRESSURELESS SLABS}

\author{Bo Li\altaffilmark{1}}
    \email{bbl@sdu.edu.cn}
\author{Ming-Zhe Guo\altaffilmark{1}}
\author{Hui Yu\altaffilmark{1}}
\and
\author{Shao-Xia Chen\altaffilmark{1}}

\altaffiltext{1}{Shandong Provincial Key Laboratory of Optical Astronomy and Solar-Terrestrial Environment,
     Institute of Space Sciences, Shandong University, Weihai 264209, China}

\begin{abstract}
Impulsively generated sausage wave trains in coronal structures are important for interpreting a substantial number
    of {  observations of} quasi-periodic signals with quasi-periods of order seconds.  
We have previously shown that the Morlet spectra of these wave trains in coronal tubes
    depend crucially on the dispersive properties of trapped sausage waves, the existence of cutoff axial wavenumbers
    and the monotonicity of the dependence of the {  axial group speed on the axial wavenumber} in particular. 
This study examines the difference a slab geometry may introduce, for which purpose 
    we conduct a comprehensive eigenmode analysis, both analytically and numerically,
    on trapped sausage modes in coronal slabs with a considerable number of density profiles.
{  
For the profile descriptions examined,
    coronal slabs can trap sausage waves with longer axial wavelengths,}   
    and {  the group speed}
    approaches the internal Alfv\'en speed more rapidly at large {  wavenumbers} in the cylindrical case. 
However, common to both geometries, cutoff wavenumbers exist only when the density profile falls sufficiently
    rapidly at distances far from coronal structures.
Likewise, the monotonicity of the {  group speed} curves depends critically on the profile steepness 
    right at the structure axis. 
Furthermore, the Morlet spectra of the wave trains are shaped by the {  group speed} curves for coronal slabs and tubes alike.
Consequently, we conclude that these spectra have the potential for telling the sub-resolution density structuring inside coronal structures,
    although their detection requires an instrumental cadence
    of better than $\sim 1$ second. 
\end{abstract}
\keywords{magnetohydrodynamics (MHD) --- Sun: flares --- Sun: corona --- Sun: magnetic fields --- waves}

\section{INTRODUCTION}
\label{sec_intro}
Magnetohydrodynamic (MHD) waves and oscillations abound in
    the highly structured solar atmosphere
    \citep[see e.g.,][for recent reviews]{2005LRSP....2....3N,2007SoPh..246....3B,2012RSPTA.370.3193D,2016GMS...216..395W,2016SSRv..200...75N}.
On the one hand, these waves may play an important role in atmospheric heating
    \citep[see e.g.,][for some recent reviews]{2006SoPh..234...41K, 2012RSPTA.370.3217P,  2015RSPTA.37340261A, 2015RSPTA.37340269D}.
On the other hand, when placed in the framework of MHD wave theory, the measurements of these waves and oscillations
    can help yield the atmospheric parameters that prove difficult to measure directly.
This practice was originally proposed for coronal applications \citep[e.g.,][]{1970PASJ...22..341U,1970A&A.....9..159R,1975IGAFS..37....3Z}
    and hence termed coronal seismology
    (\citeauthor{1984ApJ...279..857R}~\citeyear{1984ApJ...279..857R};
    also see the reviews by, e.g.,
    \citeauthor{2000SoPh..193..139R}~\citeyear{2000SoPh..193..139R},
    \citeauthor{2008IAUS..247....3R}~\citeyear{2008IAUS..247....3R}).
With the advent of advanced space-borne and ground-based instruments, MHD waves and oscillations have also been identified
    in jet-like structures such as spicules
    (e.g., \citeauthor{2007Sci...318.1574D}~\citeyear{2007Sci...318.1574D},
    \citeauthor{2009ApJ...705L.217H}~\citeyear{2009ApJ...705L.217H},
    and the review by
    \citeauthor{2009SSRv..149..355Z}~\citeyear{2009SSRv..149..355Z})
    and network jets \citep{2014Sci...346A.315T},
    prominences~\citep[e.g.,][and references therein]{2012LRSP....9....2A},
    pores and sunspots
    \citep[e.g.,][]{2008IAUS..247..351D, 2011ApJ...729L..18M, 2014A&A...563A..12D,2015ApJ...806..132G,2016ApJ...817...44F},
    as well as various chromospheric structures
    \citep[e.g.,][]{2009Sci...323.1582J,2012NatCo...3E1315M}.
For this reason, the seismological practice is now commonly referred to as solar magneto-seismology (SMS).
Furthermore, while traditionally applied to the inference of the physical parameters of localized structures,
    the ideas behind SMS have also been extended to the so-called ``global coronal seismology''
    \citep{2005A&A...435.1123W, 2007SoPh..246..177B}, which capitalizes on the measurements of various large-scale
    waves to deduce the global magnetic field in the corona
    \citep[see e.g.,][for recent reviews]{2014SoPh..289.3233L,2015LRSP...12....3W,2016GMS...216..381C}.

{  We} restrict ourselves to the applications of SMS to the solar corona. 
Perhaps the magnetic field strength in coronal loops tops the list
    of the physical parameters that one would like SMS to offer.
This is understandable because the magnetic field is known to play an essential role in shaping the corona
    but is notoriously difficult to directly measure \citep[e.g.,][]{2009SSRv..144..413C}.
Indeed, standing kink modes have largely been exploited for this purpose
    since they were first imaged by the Transition Region and Coronal Explorer
    (TRACE,
    \citeauthor{1999ApJ...520..880A}~\citeyear{1999ApJ...520..880A}
    and
    \citeauthor{1999Sci...285..862N}~\citeyear{1999Sci...285..862N})
    and subsequently by Hinode \citep{2008A&A...482L...9O, 2008A&A...489L..49E},
    the Solar TErrestrial RElations Observatories (STEREO,
    \citeauthor{2009ApJ...698..397V}~\citeyear{2009ApJ...698..397V}
    )
    and the Solar Dynamics Observatory/Atmospheric Imaging Assembly 
    (SDO/AIA, e.g., 
    \citeauthor{2011ApJ...736..102A}~\citeyear{2011ApJ...736..102A};
    \citeauthor{2012A&A...537A..49W}~\citeyear{2012A&A...537A..49W}).
It is just that the longitudinal Alfv\'en time ($L/\vai$ with $L$ being the loop length and $\vai$ being the Alfv\'en speed
    at the loop axis)
    rather than the magnetic field strength itself
    is the direct outcome of the measured kink mode periods.
However, inferring the information on density structuring transverse to coronal structures can be as important,
    because this structuring significantly affects the {  heating efficiencies} of such wave-based mechanisms
    as phase-mixing \citep{1983A&A...117..220H} and resonant absorption 
    \citep[e.g.,][]{1988JGR....93.5423H, 2002ApJ...577..475R, 2002A&A...394L..39G}.
To proceed, let $\rhoi$ ($\rhoe$) denote the density at the loop axis (in the ambient corona),
    $R$ denote the loop radius, and $l$ denote the transverse density {  length scale}.
If one attributes the damping of standing kink modes to resonant absorption,
    then their periods and damping rates can be combined to constrain
    $L/\vai$ and the transverse density structuring 
    as characterized by $\rhoi/\rhoe$ and $l/R$ \citep[e.g.,][]{2007A&A...463..333A,2008A&A...484..851G, 2014ApJ...781..111S, 2014A&A...565A..78A}.
    
Standing sausage modes in magnetized loops can also be exploited for inferring both the magnetic field strength
    and the transverse density structuring.
To see this, consider the simplest case where the gas pressure is negligible and coronal structures
   are seen as straight, field-aligned cylinders with physical parameters transversely distributed
   in a top-hat manner.
Furthermore, focus for now on the lowest-order modes.
It is well-known that trapped modes arise when the axial wavenumber $k$ exceeds some cutoff value $\kc$,
   and leaky modes arise when the opposite is true
   \citep[e.g.,][]{1983SoPh...88..179E,1986SoPh..103..277C,2007AstL...33..706K, 2012ApJ...761..134N, 2015SoPh..290.2231C, 2016ApJ...833..114C}.
It turns out that the period $P \approx 2.6 R/\vai$ and the damping-time-to-period ratio
   $\tau/P \approx (\rhoi/\rhoe)/\pi^2$ for standing sausage modes in sufficiently long loops \citep[e.g.,][]{2007AstL...33..706K}. 
This means that one can readily derive the transverse Alfv\'en time $R/\vai$ from the measured periods 
   and the density contrast $\rhoi/\rhoe$ from the measured damping-time-to-period ratios,
   provided that lateral leakage is the mechanism for the apparent wave damping. 
Things become more complicated if one considers that the transverse density distribution
   is more likely to be continuous.
In this case, $P$ and $\tau/P$ turn out to depend on the transverse density {  length scale} as well.    
This then enables one to constrain the combination $[R/\vai, \rhoi/\rhoe, l/R]$, as illustrated in 
   \citet{2015ApJ...812...22C} where we adopted the widely accepted idea that a substantial fraction of 
   quasi-periodic pulsations (QPPs) in solar flare emissions are attributable to standing sausage modes
   in flare loops \citep[e.g.,][]{2009SSRv..149..119N,2016SoPh..291.3143V}.
Let us note that the dimensionless transverse density {  length scale} $l/R$ is the most difficult to constrain,
   an issue also associated with applications of standing kink modes in active region loops \citep[e.g.][]{2007A&A...463..333A,2014ApJ...781..111S}.
However, we showed in \citet{2016SoPh..291..877G} that
   the parameter set $[R/\vai, \rhoi/\rhoe, l/R]$ can be constrained to rather narrow ranges
   by using the measurements of co-existing standing kink and sausage modes in a flare loop that
   occurred on 14 May 2013 as imaged with the Nobeyama Radio Heliograph
   (NoRH, \citeauthor{2015A&A...574A..53K}~\citeyear{2015A&A...574A..53K}). 
The reason is simply that in this case we have more knowns than when only either a kink or sausage mode 
   is measured.

Seismological applications can also be made of fast sausage wave trains in coronal structures.
Theoretically predicted by \citet{1983Natur.305..688R} 
   (see also \citeauthor{1984ApJ...279..857R}~\citeyear{1984ApJ...279..857R} 
   and \citeauthor{2015ApJ...806...56O}~\citeyear{2015ApJ...806...56O})
   and numerically examined by a series of studies
   \citep[e.g.,][]{1993SoPh..144..101M, 1994SoPh..151..305M,2004A&A...422.1067S, 2004MNRAS.349..705N, 2015ApJ...814..135S, 2016ApJ...833...51Y},
   such modes can occur as the response of coronal structures to
   impulsive, internal, localized drivers such as flaring activities.
When measured along the structure axis at a distance sufficiently far from the driver,
   the generated signals consist of three distinct phases.
A nearly monochromatic ``periodic phase'' appears first,
   then a ``quasi-periodic phase'' arises where the signals are expected to be stronger
   with quasi-periods decreasing with time,
   and finally a monochromatic ``decay or Airy phase'' occurs.
For coronal tubes with top-hat transverse density profiles as examined in \citet{1983Natur.305..688R} and \citet{1984ApJ...279..857R},
   the quasi-periods in the first phase are also approximately $2.6 R/\vai$.
On the one hand, this makes it no surprise
   that the candidates for impulsively generated sausage wave trains
   are mostly identified in high-cadence measurements in,
   say, optical passbands with ground-based instruments at total eclipses
   \citep[e.g.][]{1984SoPh...90..325P,1987SoPh..109..365P, 2001MNRAS.326..428W, 2002MNRAS.336..747W,
   2003A&A...406..709K,2016SoPh..291..155S}.
On the other hand, the measured quasi-periods can then offer an estimate of the transverse Alfv\'en time $R/\vai$
   \citep[e.g,][]{1990SoPh..130..161F, 2008IAUS..247....3R}.
   
Something particularly noteworthy in the measurements of candidate sausage wave trains
   is the identification of tadpole-shaped Morlet spectra characterized by narrow tails preceding broad heads
   \citep[e.g.,][]{2001MNRAS.326..428W,2003A&A...406..709K,2009A&A...502L..13M, 2016SoPh..291..155S}.
The appearance of these ``crazy tadpoles'' largely relies on two aspects of the
   dependence on the axial wavenumber of the group speeds of trapped sausage modes
   \citep[see][hereafter paper I]{2017ApJ...836....1Y}.
One is the existence of cutoff wavenumbers, which guarantees that the lowest-order modes
   are primarily excited provided that the impulsive driver is not too localized
   \citep[see also][]{2015ApJ...806...56O,2015ApJ...814..135S}.
The other is that with $k$ increasing from the cutoff values,
   the group speed $\vgr$ typically first decreases rapidly from the external Alfv\'en speed $\vae$ to some local minimum $\vgmin$
   before increasing towards $\vai$.
The wave packets pertinent to the portion of the $\vgr-k$ curves with $\vai < \vgr <\vae$
   account for the narrow tails in the ``crazy tadpoles'',
   because the corresponding frequencies vary little.
On the other hand, wave packets in the portion embedding $\vgmin$ can have the same group speeds but different frequencies,
   meaning that they can arrive simultaneously at the point where the measurements are made.
The superposition of multiple wavepackets makes the signals stronger, thereby accounting for
   the broad heads in the ``crazy tadpoles''.
However, these two features of the {  group speed} curves are mostly found for coronal tubes with top-hat profiles.
As shown by \citet{2015ApJ...810...87L} and paper I, cutoff wavenumbers may not exist if the density profile far from the tubes
   does not fall off sufficiently rapidly with distance.
Furthermore, both \citet{2016ApJ...833...51Y} and paper I indicated that
   the {  group speed} curves can behave in a monotonical manner
   for coronal tubes with diffuse boundaries.
It then follows that overall four different types of {  group speed} curves arise, depending on the existence of cutoff wavenumbers
   and whether the curves behave in a monotonical fashion.
Paper I demonstrated that the temporal evolution of impulsively generated wave trains in general, their Morlet spectra in particular,
   can be quite different for qualitatively different {  group speed} curves.
For instance, in the absence of wavenumber cutoffs, higher-order modes can be readily excited, resulting in the ``fins'' sitting atop
   the main bodies of the pertinent Morlet spectra.
On the other hand, when the {  group speed} curves are monotonical, the main bodies may look like obliquely directed carps rather than tadpoles.
Given that the behavior of the {  group speed} curves is determined by the transverse density distribution,
   paper I went on to suggest that this largely unknown distribution can be deduced by digging into the available data
   to look for those Morlet spectra that look drastically different from ``crazy tadpoles''.

The present study is intended to continue our paper I by examining impulsively generated sausage waves in coronal slabs.
The reason for doing this is threefold.
First, in certain circumstances, a slab geometry is more appropriate for describing the measured waves
   in the solar atmosphere
   \citep[e.g.,][]{2005A&A...430L..65V,2010ApJ...714..644C,2011ApJ...728..147C}.
In particular, impulsively generated sausage waves in a slab geometry have been invoked to account for
   a substantial number of radio emission features such as decimetric fiber bursts
   \citep[e.g.][]{2009A&A...502L..13M,2011SoPh..273..393M,2013A&A...550A...1K}
   and zebra-pattern structures in type IV radio bursts \citep{2016ApJ...826...78Y}.
Second, sausage waves in a slab geometry are easier to handle mathematically.
We will capitalize on this mathematical simplicity to offer a rather extensive analytical eigenmode analysis.
The obtained results, together with the numerical computations, will enable us to quantify
   the difference a slab geometry introduces to the dispersive properties of trapped sausage waves.  
Third, we will explicitly show that whether the {  group speed} curves are monotonical is almost entirely
   determined by the steepness of the density profile at the structure axis,
   which was implicated but not articulated in paper I.

This manuscript is organized as follows.
Section~\ref{sec_method_sol} starts with a description of the equilibrium configuration,
   and then presents the governing equations together with their solution methods.
Our methodology for studying impulsively generated wave trains is illustrated
   in Section~\ref{sec_tophat} where the much-studied top-hat profiles are examined.
A comparison for this simplest situation between the slab and cylindrical geometries 
   shows that the results are qualitatively the same.
This then inspired us to present in Section~\ref{sec_phil_compa}
   our philosophy on how to examine the quantitative differences.
Sections~\ref{sec_monomu} to \ref{sec_innermu} then examine three families of density profiles in substantial detail.
A combination of parameters that characterize sausage modes
   will be quantitatively compared with what we found in the cylindrical case.
Finally, Sect.~\ref{sec_conc} closes this manuscript with our summary
    and some concluding remarks.
To corroborate the results on the general behavior of the {  group speed}
    curves found in the main text,
    we will offer a number of appendices to examine some
    additional density profiles that permit analytical treatments.

\section{GENERAL DESCRIPTION OF METHODS FOR EXAMINING IMPULSIVELY GENERATED SAUSAGE WAVES}
\label{sec_method_sol}

\subsection{Description of the Equilibrium Slab}
\begin{figure}
\centering
\includegraphics[width=0.7\columnwidth]{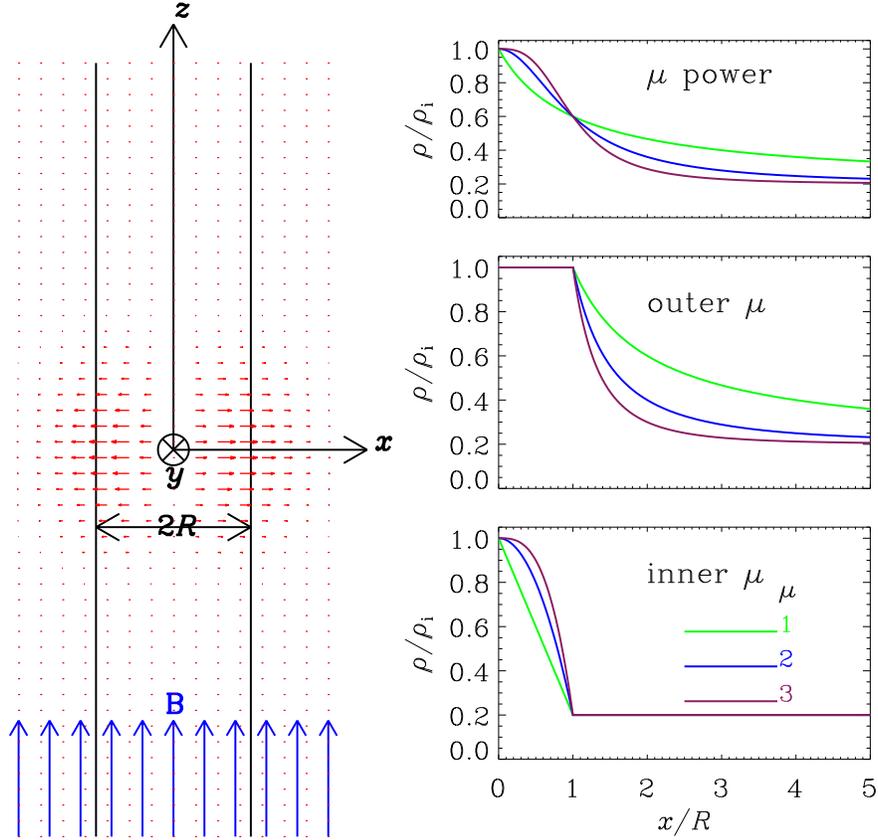}
\caption{
 Description of the modeled structured corona.
 In the left column, in addition to an illustration of the coronal slabs,
     the initial perturbation to
     the transverse velocity is also shown by the red arrows (see Equation~\ref{eq_IC_vx}).
 Shown in the right column are the three families of transverse density profiles
     examined in this study.
 For illustration purposes, the density contrast $\rho_{\rm i}/\rho_{\rm e}$ is chosen to be $5$,
     while a number of different steepness parameters ($\mu$) are chosen as labeled.
}
 \label{fig_illus_profile}
\end{figure}

From the outset, let us restrict ourselves to typical coronal applications by working in
     the framework of cold MHD, in which case slow modes are absent.
Adopting a Cartesian coordinate system $(x, y, z)$, we assume that the equilibrium magnetic field ${\bf B}$
     is uniform and directed in the $z$-direction (${\bf B} = B\hat{z}$).
To model a structured corona, we further assume that the equilibrium density $\rho$
     is a function of $x$ only and symmetric about $x=0$.
In the half-plane $x \ge 0$, it takes the form
\begin{equation}
 {\rho}(x)= \rhoe+(\rhoi-\rhoe) f(x),
    \label{eq_rho_profile_general}
\end{equation}
   where the function $f(x)$ decreases from unity at $x=0$ to zero when $x$ approaches infinity.
Let $R$ be the spatial scale that characterizes the variation of $f(x)$.
Equation~(\ref{eq_rho_profile_general}) then mimics a density-enhanced slab with half-width $R$ embedded in an ambient corona,
    with the density at the slab axis being $\rhoi$ and that far from the slab being $\rhoe$.
The Alfv\'en speed is given by $v_{\rm A} (x) = B/\sqrt{4\pi \rho (x)}$, which increases from $\vai$
   to $\vae$ with $v_{\rm A i, e} = B/\sqrt{4\pi \rho_{\rm i, e}}$.
Evidently, $\vae^2/\vai^2 = \rhoi/\rhoe$.
The equilibrium configuration is illustrated in the left column of Fig.~\ref{fig_illus_profile}.

Following Paper I, we will focus on three families of profiles.
The first one is called ``$\mu$ power'' and described by
\begin{equation}
   f(x)= \displaystyle\frac{1}{1+\left(x/R\right)^{\mu}}~.
\label{eq_profile_monomu}
\end{equation}
The next one, called ``outer $\mu$'', is given by
\begin{equation}
   f(x)=\left\{
   \begin{array}{ll}
      1,     			& 0 \le x \le R, 			\\
      \left(x/R\right)^{-\mu}, 	& x \ge R.
   \end{array} \right.
\label{eq_profile_outermu}
\end{equation}
The third one is called ``inner $\mu$'' and in the form
\begin{equation}
   f(x)=\left\{
   \begin{array}{ll}
      1-\left(\displaystyle\frac{x}{R}\right)^\mu,     	& 0 \le x \le R, \\
      0, 						& x \ge R.
   \end{array} \right.
\label{eq_profile_innermu}
\end{equation}
These profiles are illustrated in the right column of Fig.~\ref{fig_illus_profile} where
    $\rhoi/\rhoe$ is arbitrarily chosen to be $5$.
Evidently, $\mu$ is a measure of the profile steepness, and all profiles converge to a top-hat one
    when $\mu \rightarrow \infty$.
We will consider only the cases $\mu \ge 1$, because otherwise the
    ``$\mu$ power'' and ``inner $\mu$'' profiles will become cusped around $x=0$.
With typical EUV active region loops in mind, we will consider density contrasts
    in the range between $2$ and $10$ \citep[e.g.,][]{2004ApJ...600..458A}.
Note that this range is also typical of polar plumes~\citep[][Table 3]{2011A&ARv..19...35W}.
In fact, the results from this study apply to both magnetically open and closed structures.

{  All the profile prescriptions} in Equations~\eqref{eq_profile_monomu} to \eqref{eq_profile_innermu}
   have the following characteristics,
\begin{eqnarray}
f(x) \approx
   \left\{
   \begin{array}{ll}
       1-\displaystyle\left(\frac{x}{R}\right)^{\mu_0},      	& 0 < x/R \ll 1, \\
       \displaystyle\left(\frac{x}{R}\right)^{-\mu_\infty}, 	& x/R \gg 1.
   \end{array} 
   \right. 
\label{eq_def_mu0_muinfty}   
\end{eqnarray}
These characteristics, together with the specific values of $\mu_0$ and $\mu_\infty$,
   are summarized in the first four columns in Table~\ref{tbl_1}.
Evidently, with the exception of the ``$\mu$ power'' profile,
   the approximate sign in Equation~\eqref{eq_def_mu0_muinfty} is actually exact.
{  
These three families of profile prescriptions are representative in the sense that
   one can distinguish between the steepness at the slab axis 
   (characterized by $\mu_0$) and that at infinity ($\mu_\infty$).
However, obviously they do not exhaust all possible ways for prescribing the 
   transverse density distribution.
}

\subsection{Governing Equations and Methods of Solution}
To examine how the modeled equilibrium responds to small-amplitude initial disturbances,
    let $\delta \rho, \delta \vec{v}$ and $\delta \vec{b}$ represent the density,
    velocity, and magnetic field perturbations, respectively.
Neglecting propagation out of the $x-z$ plane ($\partial/\partial y \equiv 0$)
    and focusing on compressible perturbations,
    one finds from linearized cold MHD equations that
    only $\delta \rho$, $\delta v_x$, $\delta b_x$ and $\delta b_z$ survive.
A single equation can be readily derived
    for $\delta v_x(x, z; t)$, namely
\begin{equation}
\label{eq_vx_2ndorder}
 \frac{\partial^2 \delta v_x}{\partial t^2}
    =v_{\rm A}^2(x)
    \left(\displaystyle\frac{\partial^2}{\partial z^2}
         +\displaystyle\frac{\partial^2}{\partial x^2}
    \right)\delta v_x~.
\end{equation}
The density perturbation $\delta \rho$ is governed by
\begin{equation}
\displaystyle
 \frac{\partial \delta \rho}{\partial t}
    =-\left(\delta v_x\frac{\partial \rho}{\partial x}
    +\rho\frac{\partial \delta v_x}{\partial x}\right)~.
 \label{eq_rho}
\end{equation}

For each profile with a given combination of $[\rhoi/\rhoe, \mu]$, we will always start
   with an eigenmode analysis to establish the dispersive behavior of trapped sausage modes.
Fourier-decomposing any perturbation $\delta g(x,z;t)$ as
\begin{eqnarray}
\label{eq_Fourier_ansatz}
  \delta g(x,z;t)={\rm Re}\left\{\tilde{g}(x)\exp\left[-i\left(\omega t-kz\right)\right]\right\}~,
\end{eqnarray}
   one finds from Equation~(\ref{eq_vx_2ndorder}) that
\begin{eqnarray}
\label{eq_Fourier_xi}
    \frac{\mathd^2 \tilde{\xi}}{\mathd x^2}
   +\left(\frac{\omega^2}{v^2_{\rm A}}-k^2\right)\tilde{\xi}=0~,
\end{eqnarray}
    where $\tilde{\xi} = i\tilde{v}_x/\omega$ is the Fourier
    amplitude of the transverse Lagrangian displacement.
With trapped sausage modes in mind, Equation~(\ref{eq_Fourier_xi}) constitutes a standard eigenvalue problem (EVP)
     when supplemented with the following boundary conditions (BCs)
\begin{eqnarray}
  \tilde{\xi} (x = 0) = 0~, \hspace{0.2cm}
  \tilde{\xi} (x \to \infty) \to 0~.
\label{eq_BVP_BC}
\end{eqnarray}
To solve this EVP,
    we employ a MATLAB boundary-value-problem solver BVPSuite in its eigen-value mode
    (see \citeauthor{2009AIPC.1168...39K}~\citeyear{2009AIPC.1168...39K} for a description of the code,
    and see \citeauthor{2014A&A...568A..31L}~\citeyear{2014A&A...568A..31L} for an extensive validation study).
What comes out is that the dimensionless angular frequency $\omega R/\vai$ can be formally expressed as
\begin{eqnarray}
    \displaystyle \frac{\omega R}{v_{\rm Ai}} = {\cal G}\left[kR; \frac{\rhoi}{\rhoe}, f\left(\frac{x}{R}\right)\right] .
\label{eq_omega_formal}
\end{eqnarray}
Note that both $\omega$ and the axial wavenumber $k$ are real-valued.
The axial phase and group speeds simply follow from the definitions
    $\vph = \omega/k$ and $\vgr = \mathd \omega/\mathd k$, respectively.

The dispersive properties of trapped sausage modes will provide the context for interpreting
    the numerical results that examine how coronal slabs respond to a localized initial perturbation.
Instead of directly solving the time-dependent linear cold MHD equations, we choose to solve
    the equivalent version, Equation~(\ref{eq_vx_2ndorder}), which is simpler to handle.
To this end, we discretize Equation~(\ref{eq_vx_2ndorder}) in a finite-difference (FD) manner
    on a computational domain extending from $0$ to $L_x$ ($-L_z/2$ to $L_z/2$)
    in the $x$-($z$-) direction.
For simplicity, a uniform grid spacing with $\Delta z = 0.08 R$ is adopted in the $z$-direction.
However, to speed up our computations, the grid points in the $x$-direction are chosen to be
    nonuniformly distributed.
Let $\Delta x_j$ denote the spacing at grid index $j$ ($j=1, 2, 3, \cdots$).
The spacing is fixed at $\Delta x_1 = 0.02 R$ when $x \le 3 R$.
We then allow $\Delta x_j$ to increase as $\Delta x_{j+1} = 1.025 \Delta x_{j}$
    until it reaches $\sqrt{\rho_{\rm i}/\rho_{\rm e}} \Delta x_1$.
From there on $\Delta x_j$ remains a constant again.
Despite this complication, we ensure that the FD approximations to
    the spatial derivatives in Equation~(\ref{eq_vx_2ndorder}) are second order accurate
    in both the $x$- and $z$- directions.
We adopt the leap-frog method for time integration, for which a uniform time
    step $\Delta t = 0.6 \Delta_{\rm min}/v_{\rm A, max}$ is chosen to comply with
    the Courant condition.
Here $\Delta_{\rm min}$ ($v_{\rm A, max}$) represents the smallest (largest) value that
    the grid spacing (Alfv\'en speed) attains in the entire computational domain.

The boundary and initial conditions are specified as follows.
At the left boundary $x = 0$, we fix $\delta v_x$ at zero.
We choose both $L_x$ and $L_z$ to be sufficiently large such that the perturbations reflected
    off the pertinent boundaries will not contaminate the numerical results to be analyzed.
While in this regard the boundary conditions therein are irrelevant,
    we nonetheless fix $\delta v_x$ at zero in practice.
To initiate our simulations, we choose to perturb only the transverse velocity, meaning that
\begin{eqnarray}
    \frac{\partial \delta v_x}{\partial t}(x, z; t=0) = 0~,
    \label{eq_IC_vx_deriv}
\end{eqnarray}
    given the absence of $\delta b_x$ and $\delta b_z$ at $t=0$.
Finally, $\delta v_x$ at $t=0$ is specified as
\begin{eqnarray}
\displaystyle
     \frac{\delta v_x(x, z; t=0)}{\vai}
   = {\rm e}^{1/2} \left(\frac{x}{\sigma_x}\right)
     \exp\left(-\frac{x^2}{2 \sigma_x^2}\right)
     \exp\left(-\frac{z^2}{2 \sigma_z^2}\right) ,
\label{eq_IC_vx}
\end{eqnarray}
    which ensures the parity of the generated wave trains by not displacing the slab axis.
Here a constant $\exp(1/2)$ is introduced such that the right hand side (RHS)
    of Equation~(\ref{eq_IC_vx}) attains a maximum of unity.
Furthermore, $\sigma_x$ ($\sigma_z$) determines the extent
    to which the initial perturbation spans in
    the transverse (axial) direction.
Throughout this study, both $\sigma_x$ and $\sigma_z$ are chosen to be $R/\sqrt{2}$
    such that the initial perturbation
    is neither too localized nor too extended.
This perturbation is shown by the red arrows in the left column of Fig.~\ref{fig_illus_profile}.

Several points need to be made.
First, the temporal signal we actually analyze with a wavelet analysis
    is the density variation $\delta\rho$.
To find this,  
    Equation~(\ref{eq_rho}) is advanced simultaneously with Equation~(\ref{eq_vx_2ndorder})
    by assuming that $\delta\rho(x, z; t=0) = 0$. 
Second, for each profile choice we will examine the dispersive properties of trapped sausage modes
    by solving the EVP (Equations~\ref{eq_Fourier_xi} and \ref{eq_BVP_BC}) with BVPSuite.
An analytical dispersion relation (DR) will be derived when mathematically tractable.
The BVPSuite results are found to agree exactly with the numerical solutions to the pertinent DRs without exception,
    a point that we will not mention later.
We will also derive the expressions for the cutoff wavenumbers as well as the behavior
    of the group speeds $\vgr$ at large axial wavenumbers.
The latter is necessary because the asymptotic behavior of $\vgr$
    largely determines whether the $\vgr-k$ curves are monotonical.
Third, it turns out that the eigenmode analysis always results in an infinite number of branches of solutions.
We use the transverse harmonic number $l$ ($l=1, 2, 3, \cdots$) to label a branch, following the convention that
   $l=1$ refers to the fundamental transverse mode and $l\ge 2$ refers to its transverse harmonics.

\section{CORONAL SLABS WITH TOP-HAT PROFILES}
\label{sec_tophat}
Let us start with the simplest case where the transverse density profiles
    are piece-wise constant.
The reason for doing this exercise is to illustrate our methodology for examining impulsively generated wave trains.
While the group speed behavior was traditionally invoked to interpret
    the temporal evolution \citep[e.g.,][]{1993SoPh..144..101M,1994SoPh..151..305M}
    and Morlet spectra \citep[e.g.,][]{2004MNRAS.349..705N, 2012A&A...537A..46J}
    of impulsively generated wave trains in coronal slabs,
    the Morlet spectra have not been directly placed in the context of the {  group speed} curves.

\subsection{Analytical Results from the Eigenmode Analysis}
\label{sec_sub_EVP_tophat}
The following definitions are necessary throughout,
\begin{eqnarray}
\displaystyle
&&  n^2 \equiv \frac{\omega^2}{\vai^2} - k^2~, \nonumber \\
&&  m^2 \equiv k^2 - \frac{\omega^2}{\vae^2}~, \\
&&  D   \equiv n^2+m^2 = \frac{\omega^2}{\vai^2}-\frac{\omega^2}{\vae^2}~. \nonumber
\label{eq_def_nmD}
\end{eqnarray}
Both $n^2$ and $m^2$ are non-negative for trapped modes.
We further define the following dimensionless parameters,
\begin{eqnarray}
\displaystyle
&&  \bar{n} \equiv n R = kR\sqrt{\frac{\vph^2}{\vai^2}-1}~,  \nonumber \\
&&  \bar{m} \equiv m R = kR\sqrt{1-\frac{\vph^2}{\vae^2}}~,  \label{eq_def_nmD_nondimen} \\
&&  \bar{D} \equiv \bar{n}^2+\bar{m}^2 = D R^2 = \frac{\omega^2 R^2}{\vai^2}\left(1-\frac{\rhoe}{\rhoi}\right)~, \nonumber
\end{eqnarray}
   where the barred symbols are used to distinguish the dimensionless from dimensional values.
These dimensionless parameters are used only when necessary.
Without loss of generality, we assume that both $n$ and $m$ (hence $\bar{n}$ and $\bar{m}$) are non-negative.

For top-hat profiles,
    the solution to Equation~(\ref{eq_Fourier_xi}) in the uniform interior (exterior) is
    proportional to $\sin(nx)$ (${\rm e}^{-mx}$).
The dispersion relation (DR) reads \citep[e.g.,][]{1982SoPh...76..239E, 2005A&A...441..371T, 2013ApJ...767..169L}
\begin{eqnarray}
    n \cot(nR)  =- m~.
\label{eq_tophat_DR}
\end{eqnarray}
It is well-known that Equation~(\ref{eq_tophat_DR}) allows an infinite number of branches of sausage modes.
For modes of any transverse harmonic number $l$, a cutoff wavenumber exists and is given by
\begin{equation}
\displaystyle
   \kcl R = \frac{(l-1/2)\pi}{\sqrt{\rhoi/\rhoe-1}}~,
\label{eq_tophat_kc}
\end{equation}
    which simply follows from the requirement  that $m = 0$ at the cutoff.

To capitalize on the simplicity of the DR, let us further examine what happens when $k$ is in the immediate vicinity
    of its cutoff values.
For this purpose, let us define
\begin{eqnarray}
&&  k = \kcl (1+\delta_k)~,~~~ \nonumber \\
&&  \omega = \ocl (1+\delta_\omega)~, \label{eq_def_deltak_deltaomega} \\
&&  \bar{n} = \nbarcl (1+\delta_n)~, \nonumber
\end{eqnarray}
    where $0 < \delta_k, \delta_\omega, \delta_n \ll 1$, and
\begin{eqnarray}
\displaystyle
 \ocl = \kcl \vae~,~~~ \nbarcl = \kcl R \sqrt{\frac{\rhoi}{\rhoe}-1}~.
 \label{eq_def_ocl_nbarcl}
\end{eqnarray}
From the definitions of $n$ and $m$, it is readily shown that
\begin{eqnarray}
  \delta_n = \frac{(\rhoi/\rhoe) \delta_\omega-\delta_k}{\rhoi/\rhoe-1}~,~~~
  \bar{m}^2 = 2 (\kcl R)^2 (\delta_k - \delta_\omega)~.
\label{eq_def_deltan_m_2_deltakomg}
\end{eqnarray}
Note that Equations~(\ref{eq_def_deltak_deltaomega}) to (\ref{eq_def_deltan_m_2_deltakomg}) are valid for any choice of $f(x)$.
Now specialize to top-hat profiles.
The DR (\ref{eq_tophat_DR}) is equivalent to $\bar{n} = l\pi - {\rm arccot}(\bar{m}/\bar{n})$,
   which is approximately
   $\bar{n} \approx (l-1/2)\pi+\bar{m}/\bar{n}$ given that the argument $\bar{m}/\bar{n}$ is small in the present situation.
One then finds that $\bar{m} \approx \nbarcl^2 \delta_n$.
With the aid of Equation~(\ref{eq_def_deltan_m_2_deltakomg}), one finds that
\begin{eqnarray}
\displaystyle
  \delta_\omega \approx \delta_k - \frac{(\rhoi/\rhoe-1)\eta^2}{2}\delta_k^2~,
\label{eq_link_deltaomg_deltak}
\end{eqnarray}
   where $\eta=(l-1/2)\pi$.
The definitions of $\vph$ and $\vgr$ yield that $\vph = \vae(1+\delta_\omega)/(1+\delta_k)$
   and $\vgr = \vae (\mathd \delta_\omega/\mathd \delta_k)$.
One eventually arrives at
\begin{eqnarray}
\displaystyle
  \frac{\vph}{\vae} \approx 1-\frac{(\rhoi/\rhoe-1)\eta^2}{2}\left(\frac{k}{\kcl}-1\right)^2~,
\label{eq_tophat_vph_kcvic}
\end{eqnarray}
   and
\begin{eqnarray}
 \displaystyle
 \frac{\vgr}{\vae} \approx 1- \left(\rhoi/\rhoe-1\right)\eta^2\left(\frac{k}{\kcl}-1\right)~,
   \label{eq_tophat_vgr_kcvicinity}
\end{eqnarray}
    where $\delta_k$ is replaced with $k/\kcl-1$ to make the equations more self-contained.
This explicitly shows that $\vgr$ decreases from $\vae$ when $k$ increases from its cutoff values,
    even though this behavior is well-known in numerical solutions to the DR \citep[e.g., Fig.~2 in][hereafter ER88]{1988A&A...192..343E}.
In fact, it is not straightforward to anticipate that both $\vph$ and $\vgr$ attain the same value of $\vae$ at the cutoff
    because they are defined differently.
Apart from top-hat profiles, we find that the approximate behavior of $\vph$ and $\vgr$ {  with} $k$ in the neighborhood of its cutoff values
    can also be found for the exponential as well as symmetric Epstein profiles, both given in the appendix.
They are also in the same form as Equations~(\ref{eq_tophat_vph_kcvic}) and (\ref{eq_tophat_vgr_kcvicinity}).
While $\eta$ is different, it remains a function of $l$ only.

Now consider what happens when $kR \rightarrow \infty$.
One finds that $m/n \rightarrow \infty$ in this case, meaning that $\tan(nR) \rightarrow 0$.
Consequently, one finds that \citep[e.g.,][]{2013ApJ...767..169L}
\begin{equation}
\displaystyle
   \frac{\vph^2}{\vai^2} \approx 1+\left(\frac{l\pi}{k R}\right)^2~,
\label{eq_tophat_vphBigK}
\end{equation}
   and
\begin{eqnarray}
\displaystyle
    \frac{\vgr^2}{\vai^2} \approx 1-\left(\frac{l\pi}{k R}\right)^2~.
\label{eq_tophat_vgBigK}
\end{eqnarray}
This means that $\vph$ ($\vgr$) should eventually approach $\vai$ from above (below) when $kR$ increases.
\footnote{
Equation~(\ref{eq_tophat_vphBigK}) suffices for our purpose here, even though
   a more accurate approximation can be readily derived.
It reads $\vph^2/\vai^2 \approx 1+\delta^2$, where
\begin{eqnarray*}
\displaystyle
 \delta = \frac{l \pi}{kR}\left[1-\frac{1}{\sqrt{1-\rhoe/\rhoi}(kR)}\right].
\end{eqnarray*}
To arrive at this expression, we note that Equation~(\ref{eq_tophat_DR})
    is equivalent to $\tan(nR) = -n/m$, meaning that $nR = l\pi-\arctan(n/m)$.
Evidently, $\delta \ll 1$ and $n/m \ll 1$ when $kR$ is large.
Retaining only terms of order $\delta$ in the Taylor expansion of $\arctan(n/m)$,
   one sees that $\delta (kR) \approx l\pi - \delta/\sqrt{1-\rhoe/\rhoi}$.
Plugging $\delta \approx l\pi/(kR)$, the solution accurate to first order in $1/(kR)$,
   into the right hand side then yields the expression for $\delta$.
}

\subsection{Group Speed Curves}
\begin{figure}
\centering
\includegraphics[width=0.8\columnwidth]{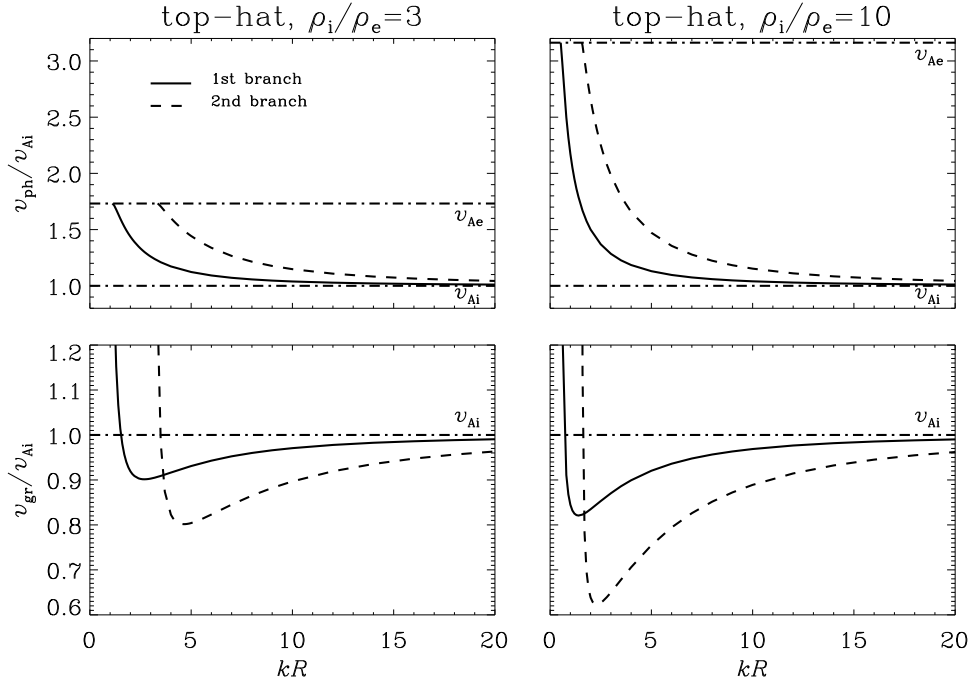}
 \caption{
 Dependence on the axial wavenumber $k$ of the axial phase (the upper row) and  group (lower) speeds
     for top-hat profiles with a density contrast of $3$ (the left column) and $10$ (right).
 The solid (dashed) curves represent the first (second) branch of trapped modes, corresponding to
     a transverse order of $1$ ($2$).
 The horizontal dash-dotted lines represent the internal and external Alfv\'en speeds ($\vai$ and $\vae$).
 {\bf Note that in the lower row, the horizontal line representing $\vae$ is absent because $\vae$ is beyond
     the range of the vertical axis.}
}
 \label{fig_vphvg_k_tophat}
\end{figure}

Figure~\ref{fig_vphvg_k_tophat} shows the dependence on the axial wavenumber $k$
    of the axial phase (the upper row) and group (lower) speeds
    for both a density contrast of $3$ (the left column) and $10$ (right).
The solid and dashed curves correspond to the fundamental transverse mode (with $l=1$)
    and its first harmonic (with $l=2$), respectively.
Furthermore, the horizontal dash-dotted lines represent the internal and external
    Alfv\'en speeds, namely $\vai$ and $\vae$.
Regarding cutoff wavenumbers, one sees that they exist for both density contrasts
    and for both branches.
In fact, Equation~(\ref{eq_tophat_kc}) suggests that this is true for arbitrary $l$
    and $\rhoi/\rhoe$ as long as $\rhoi/\rhoe >1$.
On top of that, $\kcl$ increases with $l$ but decreases with $\rhoi/\rhoe$,
    which is also expected from Equation~(\ref{eq_tophat_kc}).
Examining the lower row, one finds that
    the dependence of $\vgr$ on $k$ in the starting portion of the $\vgr-k$ curves
    is steeper when $l$ or $\rhoi/\rhoe$ increases.
This is readily understandable given the approximate behavior of $\vgr$
    as described by Equation~(\ref{eq_tophat_vgr_kcvicinity}).
In addition, all of the $\vgr-k$ curves show a non-monotonical dependence on $k$
    in that $\vgr$ first sharps decreases with $k$ before eventually increasing towards $\vai$.
This behavior can be partly understood with Equation~(\ref{eq_tophat_vgBigK}).

\subsection{Temporal Evolution and Morlet Spectra of Density Perturbations}

\begin{figure}
\centering
\includegraphics[height=80mm]{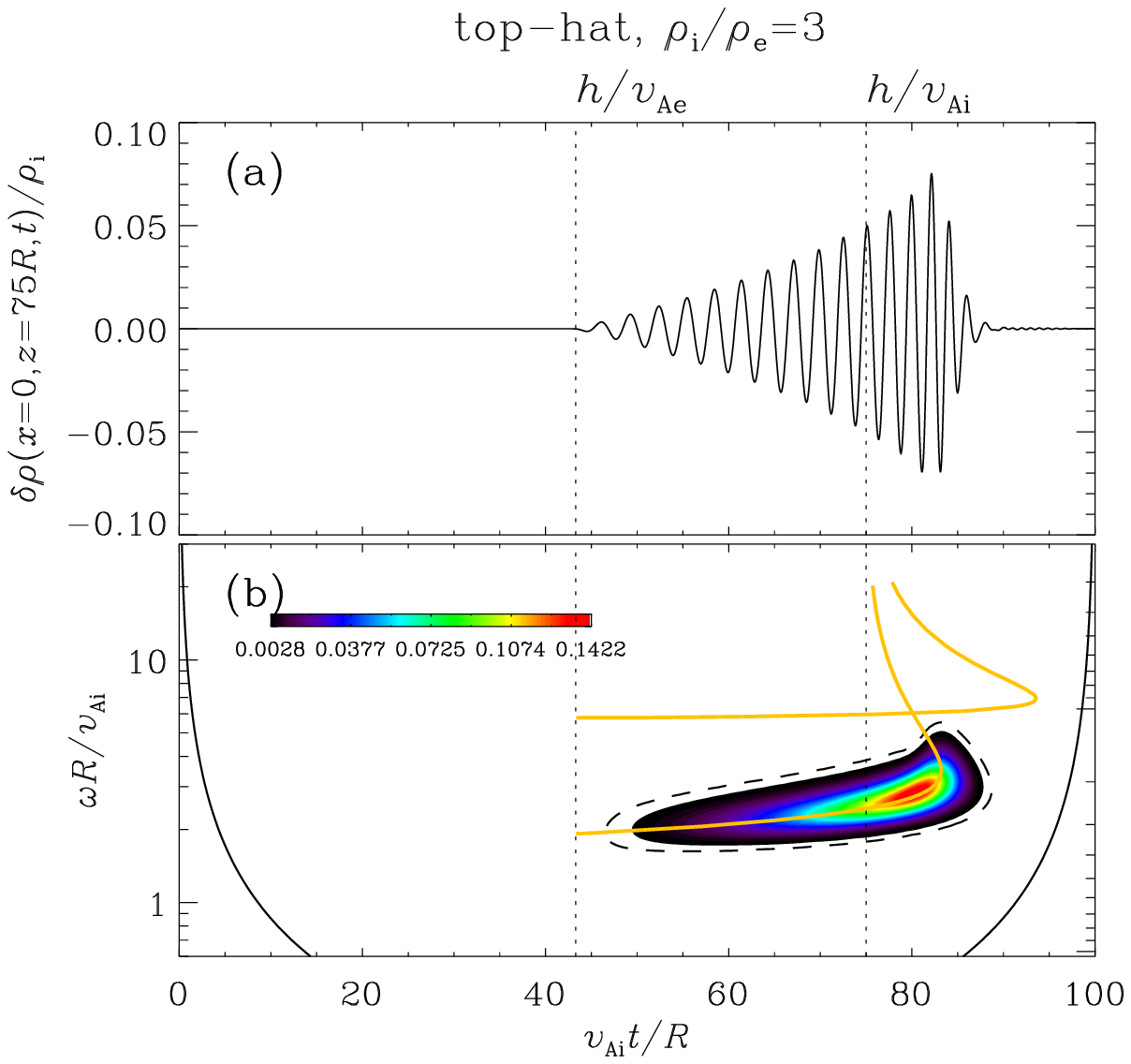}
\includegraphics[height=80mm]{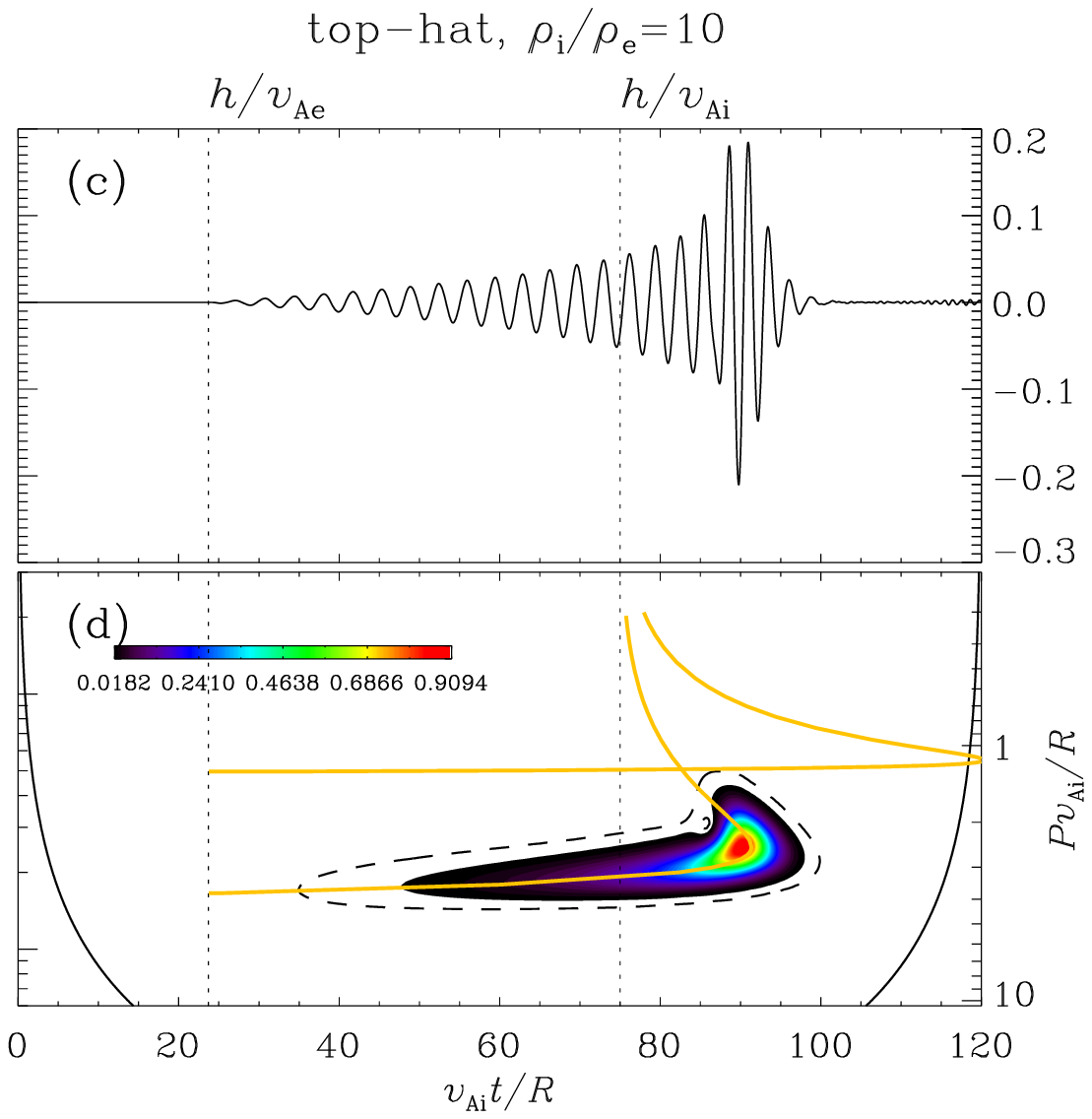}
\caption{
 Density perturbations $\delta\rho$ at a distance $h=75R$
     from the impulsive source along the axis of a coronal slab
     with top-hat profiles.
 The left and right columns pertain to a density contrast $\rho_{\rm i}/\rho_{\rm e}$
     of $3$ and $10$, respectively.
 In addition to the temporal evolution (the upper row),
     the corresponding Morlet spectra are also shown (lower).
 The left and right vertical axes in the lower row represent
     the angular frequency $\omega$ and period $P$, respectively.
 Furthermore, the black solid curves represent
     the cone of influence,
     and the area inside the dashed contour indicates where the Morlet power
     exceeds the $95\%$ confidence level.
 The dotted vertical lines correspond to the arrival times
     of wavepackets traveling at the internal and external Alfv\'en speeds as labeled.
 The yellow curves represent $\omega-h/\vgr$ as found from the eigenmode analysis,
     with the transverse order increasing from bottom to top.
}
\label{fig_wavelet_tophat}
\end{figure}

Figure~\ref{fig_wavelet_tophat} displays the temporal evolution (the upper row)
    and the pertinent Morlet spectra (lower) of
    the density perturbations $\delta\rho$ sampled at a distance $h=75R$
    along the slab axis for both a density contrast $\rhoi/\rhoe = 3$ (the left column)
    and $10$ (right).
In the lower row, the left (right) vertical axis represents the angular frequency $\omega$ (the period $P$).
The Morlet spectra are created by using the standard wavelet toolkit devised by
    \citet{1998BAMS...79...61T}, and
    the dashed contours represent the $95\%$ confidence level
    computed by assuming a white-noise process for the mean background spectrum.
The dotted vertical lines correspond to the arrival times of wavepackets
    traveling at the internal and external Alfv\'en speeds, i.e., $\vai$ and $\vae$.
The yellow curves represent $\omega$ as a function of $h/\vgr$,
    replotted with the numerical results already given in
    the lower row of Fig.~\ref{fig_vphvg_k_tophat}.
For these $\omega-h/\vgr$ curves, the corresponding transverse harmonic number
    increases from bottom to top.

Consider the lower row first.
The first impression is that the Morlet spectra are well organized by the group speed curves.
In particular, it is clear that with the present choice of the initial perturbation,
    wavepackets corresponding to the fundamental transverse mode (with $l=1$)
    dominate the signals.
Figure~\ref{fig_wavelet_tophat}d shows the typical shape of ``crazy tadpoles'', for which
    the broad head corresponds to the portion of the $\vgr-k$ curve where
    the local minimum is embedded.
This indicates that wavepackets corresponding to that portion can receive a substantial fraction
    of the energy contained in the initial perturbation, and the subsequent superposition
    of multiple wavepackets with the same group speed but different frequencies
    can account for both the enhancement of the Morlet power
    and the broadening in frequency coverage.
Note that this is in close agreement with the heuristic reasoning by \citet{1986NASCP2449..347E}.
The Morlet spectrum in Fig.~\ref{fig_wavelet_tophat}b, albeit also looking like a ``crazy tadpole'',
    is somehow different in that
    the strongest power does not enclose the group speed minimum.
The initial perturbation can still distribute a certain amount of energy to
    wavepackets beyond the group speed minimum, but this fraction is less significant
    and the superposition of wavepackets is not as clear as in Fig.~\ref{fig_wavelet_tophat}d.
Nonetheless, this fraction is still substantial enough to show up, resulting in the broadening in frequency coverage
    and the consequent appearance of a broad head.
Given that a fixed initial perturbation is adopted in this present study,
    the difference between Figs.~\ref{fig_wavelet_tophat}b and \ref{fig_wavelet_tophat}d
    indicates the importance of the density contrast in determining how the energy contained
    in the initial perturbation is distributed to different frequency ranges
    along the {  group speed} curves.

The observational implications of Figs.~\ref{fig_wavelet_tophat}b and \ref{fig_wavelet_tophat}d
    are as follows.
First, the $\omega-h/\vgr$ curves thread the narrow tails rather than bordering them from below.
As a consequence, in principle the longest period $P_{\rm Max}$ that can be resolved in the crazy tadpoles
    cannot be attributed to
    $\Pc$, the longest period that trapped modes can theoretically attain.
However, $P_{\rm Max}$ differs from $\Pc$ only marginally.
Given the practical uncertainties in determining $P_{\rm Max}$ {  (the longest period)},
    it should be fine if one equates $\Pc$ {  (the cutoff period in the trapped regime)} to $P_{\rm Max}$ in the crazy tadpoles
    as found in, say, optical observations at total eclipses~\citep[e.g.,][]{2001MNRAS.326..428W,2002MNRAS.336..747W}.
Second, the timescales associated with the strongest power do not differ much from $\Pc$ and amount to
    a couple of the transverse Alfv\'en times.
Note that {  the cutoff period} $\Pc = 2\pi/(k_{\rm c,1} \vae) = 4 (R/\vai)\sqrt{1-\rhoe/\rhoi}$ for top-hat profiles,
    as indicated by Equation~(\ref{eq_tophat_kc}).
This means that the present crazy tadpoles can be most readily found in high-cadence measurements
    made in, say, radio passbands.

\section{PHILOSOPHY FOR A COMPARATIVE STUDY BETWEEN THE SLAB AND CYLINDRICAL GEOMETRIES}
\label{sec_phil_compa}

Comparing Figure~\ref{fig_wavelet_tophat} with Figure~3 in paper I indicates that they are remarkably similar.
In fact, repeating the computations presented in paper I for the slab geometry,
   we find that the same can be said for all the Morlet spectra that are obtained.
Regardless of geometry, the Morlet spectra are all shaped by the $\omega-h/\vgr$ curves,
   which in turn derive from the frequency dependence of the axial group speed of
   trapped sausage modes.
In this regard, there is no point to present the Morlet spectra again:
   the slab results can be well anticipated with their cylindrical counterparts
   as long as we can quantify the difference that a slab geometry introduces
   to the {  group speed} curves.
Therefore, for the group speed curves and Morlet spectra pertinent to the slab geometry,
   we refer the readers to the relevant figures in paper I as listed in the last two
   columns in Table~\ref{tbl_1}.
In what follows we examine the differences between the two geometries 
    of the parameters that characterize the {  group speed} curves.

\subsection{Comparison of Cutoff Wavenumbers}

Common to both geometries, with the help of Kneser's oscillation theorem,
    one can analytically establish that cutoff wavenumbers ($\kcl$) exist
    only when the transverse density profile drops sufficiently rapidly
    at large distances.
This was shown by \citet{2015ApJ...810...87L} for the cylindrical case (also see paper I for numerical demonstrations), 
    and by \citet[][hereafter LN15]{2015ApJ...801...23L} for the slab geometry.
If $f(x)$ tends to zero in the way described by    
    Equation~\eqref{eq_def_mu0_muinfty}, 
    then this translates into that $\kcl$ exists only when $\mu_\infty \ge 2$.
Our numerical results with BVPSuite agree with this analytical expectation.
In fact, we can further demonstrate that when cutoff wavenumbers exist, they should be of the form
\begin{eqnarray}
\displaystyle
 \kcl R = \frac{d_l}{\sqrt{\rhoi/\rhoe-1}}~,
 \label{eq_kcl_general}
\end{eqnarray}
    where $d_l$ 
{  is a dimensionless parameter that measures how well a slab can trap sausage modes 
     in terms of their axial wavelengths.
It can be readily shown that $d_l$}    
    possesses no dependence on the density contrast $\rhoi/\rhoe$.
To see this, we note that the terms in the parentheses in Equation~(\ref{eq_Fourier_xi}) can be reformulated as
   $k^2 (\vph^2/v_{\rm A}^2-1)$, which reads $\kcl^2 (\rho/\rhoe-1)$ at cutoffs given that $\vph = \vae$.
With $\rho(x)$ given by Equation~(\ref{eq_rho_profile_general}),
   this then yields $\kcl^2(\rhoi/\rhoe-1)f(x)$.
As a result, $\kcl\sqrt{\rhoi/\rhoe-1}$ does not depend on the density contrast any more
   but is solely determined by $f(x)$.
Repeating the same practice for Equation~(8) in paper I, one sees that Equation~\eqref{eq_kcl_general}
   also holds for the cylindrical geometry. 
Therefore, common to both geometries, 
   Equation~(\ref{eq_kcl_general}) suggests that
    $\kcl$ always decreases with $\rhoi/\rhoe$ for a given $l$ for any profile
    describable by Equation~\eqref{eq_rho_profile_general}.   
    
The specific values of $d_l$, however, are geometry dependent.
Take the top-hat profile, which is an exemplary realization for $\mu_\infty = \infty$.
Equation~\eqref{eq_tophat_kc} indicates that $d_l = (l-1/2)\pi$ for a slab geometry
    whereas $d_l = j_{0, l}$ for a cylindrical one (see, e.g., Equation~15 in paper I).
Here $j_{0, l}$ is the $l$-th zero of $J_0$ ($l=1, 2, 3, \cdots$).
Listed in the second row of Table~\ref{tbl_2} are some specific values 
     of $d_l$ for the first several transverse harmonic numbers.
One sees that $d_l$ is always larger for a given $l$ in the cylindrical case, which is understandable
     given that $j_{0,l}$ can be well approximated by $(l-1/4)\pi$
     (see Equation~9.2.1 in \citeauthor{1972hmfw.book.....A}~\citeyear{1972hmfw.book.....A}, hereafter AS). 
Table~\ref{tbl_2} also compares another two profiles, for which $d_l$ can be analytically found in the slab geometry.
{  These two profiles, examined in Appendices \ref{sec_app_EVP_exp} and \ref{sec_app_EVP_eps},}
     are not analytically tractable in the cylindrical geometry
     and were not examined in paper I.
The pertinent values for $d_l$ are found with BVPSuite.

It is also possible to offer some rather generic analysis for 
    the cutoff wavenumbers when $\mu_\infty = 2$.
In the cylindrical case, \citet{2015ApJ...810...87L} showed that 
    $d_l = 1$.
Closely following the approach therein, we now examine what happens in the slab geometry.
By noting that $m = 0$ {  at the cutoff wavenumber},
    one finds that Equation~(\ref{eq_Fourier_xi}) at {  large distances}
    is approximately
\begin{eqnarray}
    \frac{\mathd^2 \tilde{\xi}}{\mathd x^2}
   +\frac{D R^2}{x^2}\tilde{\xi}=0~.
\end{eqnarray}
The solution to this equation is in the form
\begin{eqnarray}
\displaystyle
   \tilde{\xi} \propto \left(x/R\right)^{1/2} \sin\left[\left(\ln \frac{x}{R}\right)\sqrt{D R^2 - \frac{1}{4}}  \right] .
\end{eqnarray}
For waves to be trapped, $\tilde{\xi}$ should be non-oscillatory when $x/R \rightarrow \infty$, meaning that
    $DR^2 - 1/4 = 0$.
Given that now $DR^2 = \kcl^2 R^2 (\rhoi/\rhoe -1)$, one finds
\begin{equation}
    \kcl R = \displaystyle \frac{1/2}{\sqrt{\rhoi/\rhoe-1}}~.
    \label{eq_monomu2_DR_cutoff}
\end{equation}
In other words, $d_l = 1/2$, which does not depend on $l$.
Furthermore, the derivation shown above 
    indicates that $d_l$ does not depend on the details of $f(x)$ either.
It always reads $1/2$ as long as $f(x) \approx (x/R)^{-2}$ at large $x$.    
Therefore, it is not surprising to see that Equation~\eqref{eq_monomu2_DR_cutoff}
    was shown by LN15 to hold for an $f(x)$ being $1/(1+x/R)^2$,
    in which case an analytical DR can be found (see also appendix~\ref{sec_app_EVP_invSqr}).

The profiles examined so far suggest that $d_l$ for arbitrary $l$ is larger in the cylindrical case.
Physically speaking, this means that 
    {  magnetic slabs can trap sausage waves with longer axial wavelengths
    than magnetic cylinders}.
\footnote{This is not to be confused with the trapping capabilities in terms of energy confinement.
Take the top-hat profiles for instance.
For coronal slabs, the external transverse displacement drops off with distance as $\exp(-mx)$.
This drop-off rate is actually less rapid than in the cylindrical case, for which 
    the displacement behaves as $K_1 (mr)$~\citep[e.g.,][Equation 12]{2015ApJ...812...22C}.
For further discussions on this aspect, please see e.g.,
    \citet{2007SoPh..246..213A} and \citet{2014A&A...567A..24H}.
}
When $f(x) \approx (x/R)^{-2}$, we have shown that
    this is true regardless of the details of $f(x)$ close to the structure.
However, it remains to be examined as to how $d_l$ behaves for other cases.
Specializing to the ``inner $\mu$'' profile, for instance,
    we still need to find out how $d_l$ in the slab case differs from its cylindrical counterpart
    when $\mu_0$ varies (see Table~\ref{tbl_1}).

\subsection{Comparison of the Behavior of the Group Speed Curves at Large Wavenumbers}

For both geometries and for all the profiles given
    by Equations~\eqref{eq_profile_monomu} to  \eqref{eq_profile_innermu},
    we find that the axial phase speeds at large $kR$ take the form
\begin{equation}
  \displaystyle\frac{\vph^2}{\vai^2}
      \approx  1+\displaystyle\left(\frac{c_l}{kR}\right)^\beta .
  \label{eq_monomu_vphbigK}
\end{equation}
The exponent $\beta$ is geometry independent, and is 
     entirely determined by the steepness of $f$ at the structure axis.
In terms of $\mu_0$ (see Equation~\ref{eq_def_mu0_muinfty}), 
     we find that 
\begin{eqnarray}
\displaystyle 
   \beta = \frac{2\mu_0}{\mu_0+2} .
\label{eq_def_beta}   
\end{eqnarray}
We further find that the constant $c_l$ is of the form
\begin{eqnarray}
\displaystyle 
   c_l = h_l \left(1-\frac{\rhoe}{\rhoi}\right)^{1/\mu_0}~.
\label{eq_def_hl}   
\end{eqnarray}
    where $h_l$ does not depend on $\rhoi/\rhoe$ but is geometry dependent.
Equation~\eqref{eq_monomu_vphbigK} was found through
    an extensive parameter study with BVPSuite, and was also given in paper I
    for the cylindrical geometry.
However, it was not mentioned therein that $c_l$ takes the form given by
    Equation~\eqref{eq_def_hl}.
Actually, this apparently involved form is inspired by the analytical analysis 
    enabled by the mathematical simplicity for the slab geometry.
For the top-hat profile pertinent to $\mu_0 = \infty$,
    Equation~\eqref{eq_tophat_vphBigK} indicates that $c_l = l\pi$.
There is no need to distinguish between $\mu$ and $\mu_0$ for the ``inner $\mu$'' profile.
In this case, for $\mu = 2$ and $\mu = 1$,
    Section~\ref{sec_sub_EVP_innermu} indicates that 
    {  the dependence of $c_l$ on the density contrast 
     reads $(1-\rhoe/\rhoi)^{1/2}$
    and $(1-\rhoe/\rhoi)$}, respectively.
       
The importance of Equation~\eqref{eq_monomu_vphbigK}
    is that it largely determines whether the $\vgr-k$ curves are monotonical.
This is because the axial {  group speed} at large $kR$ is given by
\begin{eqnarray}
\displaystyle\frac{\vgr^2}{\vai^2}
    \approx
    1+\left(1-\beta\right)\left(\frac{c_l}{kR}\right)^\beta~.
\label{eq_monomu_vgrbigK}
\end{eqnarray}
It follows from Equation~\eqref{eq_def_beta} that 
    $\beta >1$ ($\beta <1$) when $\mu_0 >2$ ($\mu_0 <2$), meaning that
    $\vgr$ approaches $\vai$ from above (below) asymptotically.
Now that $\vgr$ always decreases first with increasing $k$,
    this means that the $\vgr-k$ curve is definitely nonmonotonical
    when $\mu_0 >2$ and is very likely to be monotonical 
    when $\mu_0 <2$.
    
Given the importance of Equation~\eqref{eq_monomu_vphbigK},
    we have also examined another two profiles that 
    can be approximated by $f(x) \approx 1-(x/R)^{\mu_0}$ when $x \ll R$.
Appendix~\ref{sec_app_EVP_exp} examines an $f(x)$ being 
    $\exp(-x/R)$, pertinent to $\mu_0 = 1$.
The behavior of $\vph$ at large $kR$ is in exact agreement 
    with what we found for the ``inner $\mu$'' profile with $\mu=1$.
In Appendix~\ref{sec_app_EVP_eps},     
    $f(x)$ is described by $\sech^2(x/R)$, pertinent to $\mu_0 = 2$.
The asymptotic behavior of $\vph$ is found to agree exactly with 
    the results for the ``inner $\mu$'' profile with $\mu=2$.
These two profiles are distinct from those in the main text in that
    they do not follow the form given by Equation~\eqref{eq_def_mu0_muinfty}
    at large $x$.
This means that the validity of Equation~\eqref{eq_monomu_vphbigK} 
    does not {  depend on} the details of $f(x)$ away from the slab axis.
In fact, the same can be said also for the cylindrical case.
Examining coronal tubes with these two profiles, albeit now numerically with BVPSuite,
    we find that Equations~\eqref{eq_monomu_vphbigK} to \eqref{eq_def_hl} also hold.
Furthermore, the comparison between the two geometries for these two profiles
    indicate that $h_l$ is geometry dependent (see Table~\ref{tbl_3}).

All the above-mentioned results enable us to conjecture that
\begin{conj}
  {  The phase speeds for sausage waves at large axial wavenumbers
           can be approximated by Equation~\eqref{eq_monomu_vphbigK}
           for any $f(x)$ that is approximately $1-(x/R)^{\mu_0}$ when $x/R \ll 1$.  }  
  Here $\beta$ and $c_l$ are given by Equations~\eqref{eq_def_beta} and \eqref{eq_def_hl}, respectively.
  Equation~\eqref{eq_monomu_vphbigK} is valid for both the slab and cylindrical geometries,
      barring the trivial difference that $x$ should be interpreted as the radial distance
      from the tube axis in the latter.
  Note, however, that $h_l$ is geometry dependent.    
\label{conj_vph_bigK}
\end{conj}

Conjecture~\ref{conj_vph_bigK} is supported by all the analytical results
    in both paper I and this study.
However, while mathematically simpler,
    compact closed-form DRs can be found for trapped sausage waves
    only for a handful of density profiles even in the slab geometry.
Therefore in what follows we will first show that Conjecture~\ref{conj_vph_bigK}
    is also supported by all the numerical results for profiles 
    given by Equations~\eqref{eq_profile_monomu} to \eqref{eq_profile_innermu}.
By doing this, we will also be able to address the difference in the asymptotic behavior
    of the axial phase speed in the two geometries.
For this purpose, only $h_l$ needs to be compared.

\section{CORONAL SLABS WITH ``$\mu$ POWER'' PROFILES}
\label{sec_monomu}
This section examines the ``$\mu$ power'' profiles as given by Equation~(\ref{eq_profile_monomu}).
We will start with an examination of the particular case
    {with $\mu=1$, for which an analytical treatment is possible.}

\subsection{Analytical Results for $\mu=1$}
\label{sec_sub_EVP_monomu}

The case with $\mu=1$ has already been examined by LN15.
The solution to Equation~(\ref{eq_Fourier_xi}) can be shown to have the form
\begin{eqnarray*}
     \tilde{\xi} \propto W_{\nu, 1/2}(X),
\end{eqnarray*}
    where $X= 2 m (x+R)$, and $W_{\cdot, \cdot}(\cdot)$ is Whittaker's W function
    (see section 13.1 in AS).
   \footnote{
   Another independent solution is $M_{\nu, 1/2}(X)$ with $M_{\cdot, \cdot}(\cdot)$ being Whittaker's M function.
   However, it diverges when $x$ and hence $X$ approach infinity.
   As found by LN15, $\tilde{\xi}$ can also be expressed in terms of Kummer's U function.
   However, we find that using Whittaker's W function slightly simplifies the analytical manipulations.}
In addition,
\begin{eqnarray}
   \nu = \displaystyle\frac{\bar{D}}{2\bar{m}}~.
   \label{eq_monomu1_def_nu}
\end{eqnarray}
The DR is given simply by the requirement that $\tilde{\xi}(x=0) = 0$, resulting in
\begin{eqnarray}
   W_{\nu, 1/2}(2mR) = 0~.
\label{eq_DR_monomu1}
\end{eqnarray}

No cutoff wavenumber exists for sausage modes of any transverse harmonic number $l$, given that
    $f(x)$ decreases less rapidly than $x^{-2}$ at large $x$.
When $kR \rightarrow 0$, it turns out that $\nu\rightarrow l$.
The definition of $\nu$ then yields an equation quadratic in $\omega^2$.
Solving this equation for $\omega^2$, one finds that
\begin{equation}
\displaystyle
   \frac{\vph^2}{\vae^2}
   \approx 1-\left(\frac{\rhoi/\rhoe-1}{2l}\right)^2 \left(kR\right)^2~,
   \label{eq_monomu1_vph_smallK}
\end{equation}
   and
\begin{equation}
\displaystyle
   \frac{\vgr^2}{\vae^2}
   \approx 1-3\left(\frac{\rhoi/\rhoe-1}{2l}\right)^2 \left(kR\right)^2~.
   \label{eq_monomu1_vgr_smallK}
\end{equation}
Note that LN15 also considered this situation, and an expression for $\vph$ at small $kR$
   was given by Equation~(49) therein.
However, a typo is present there in that
   $1/\chi$ (or $\rhoi/\rhoe$ with our notations)
   should be replaced by $\rhoi/\rhoe-1$.

The analytical treatment up to this point was already done by LN15.
Let us offer some new results by finding the asymptotic expressions for $\vph$ and $\vgr$ when $kR \rightarrow \infty$.
To start, let us note that $\vph$ approaches $\vai$ from above and both $mR$ and $\nu$ approach infinity.
Let $\chi$ denote $mR/(2\nu) = m^2 R^2/D$.
It is easy to show that $\chi$ approaches unity
   from below, namely $\chi \rightarrow 1^-$.
Now to evaluate the DR (\ref{eq_DR_monomu1}),
   one needs the asymptotic expansion for $W_{\nu, 1/2}(4\nu\chi)$ at large $\nu$,
   which was given on page 412 in \citet{1997asymp.book.....O} and
   is too lengthy to be included here.
Fortunately, for our purpose $W_{\nu, 1/2}(4\nu\chi)$ is dominated by a term proportional to ${\rm Ai}[(4\nu)^{2/3}\zeta]$ where
   ${\rm Ai}$ is Airy's function, and $\zeta$ denotes the solution to the following equation
\begin{eqnarray}
 \displaystyle\frac{4}{3}\left(-\zeta\right)^{3/2} =
    \arccos\left(\chi^{1/2}\right)-\left(\chi-\chi^2\right)^{1/2}~.
 \label{eq_def_zeta}
\end{eqnarray}
This means that the solution to the DR can be approximated by
\begin{eqnarray}
   \displaystyle (4\nu)^{2/3}\zeta \approx a_l \approx -\left[\frac{3\left(4l-1\right)\pi}{8}\right]^{2/3}~,
   \label{eq_def_AiZero}
\end{eqnarray}
   where $a_l$ denotes the $l$-th zero of ${\rm Ai}$ and the second approximation is accurate to within $2\%$
   (see AS, page 450).
Using the definition of $\nu$, one finds that
\begin{eqnarray}
   \displaystyle\zeta \approx a_l (4\nu)^{-2/3} = a_l \left(2kR \sqrt{1-\frac{\rhoe}{\rhoi}} \right)^{-2/3}~.
   \label{eq_monomu_approx_zeta}
\end{eqnarray}
Now let $\chi = 1-\delta$ where $0 < \delta \ll 1$.
Taylor-expanding the RHS of Equation~(\ref{eq_def_zeta}) and keeping terms up to $\delta^{3/2}$,
    one finds that $\delta = -2^{2/3}\zeta$.
Furthermore, let $(\vph/\vai)^2 = 1+\gamma$ where $\gamma$ is small and positive.
Plugging $\vph^2$ into the definitions of $m$ and $\nu$, one finds that $\chi = 1-\gamma/(1-\rhoe/\rhoi)$, meaning
    that $\gamma = (1-\rhoe/\rhoi)\delta = -(1-\rhoe/\rhoi) 2^{2/3}\zeta$.
Given Equations~(\ref{eq_def_AiZero}) and (\ref{eq_monomu_approx_zeta}), one then finds that
\begin{eqnarray}
    \displaystyle
    \frac{\vph^2}{\vai^2}
    \approx 1+\left[\frac{3\left(4l-1\right)\pi(1-\rhoe/\rhoi)}{8 kR}\right]^{2/3}~,
    \label{eq_monomu1_vphBigK}
\end{eqnarray}
    and
\begin{eqnarray}
    \displaystyle
    \frac{\vgr^2}{\vai^2}
    \approx 1+\frac{1}{3}\left[\frac{3\left(4l-1\right)\pi(1-\rhoe/\rhoi)}{8 kR}\right]^{2/3}~.
    \label{eq_monomu1_vgBigK}
\end{eqnarray}
This means that both $\vph$ and $\vgr$ eventually approach $\vai$ from above when $kR$ increases.

\subsection{Comparison with the Cylindrical Results}

Let us start by noting that 
    there is no need to distinguish between $\mu$ and $\mu_0$ or $\mu$ and $\mu_\infty$
    for this family of profiles.
In other words, $\mu_0 = \mu$ and $\mu_\infty = \mu$.    
    
\begin{figure}
\centering
\includegraphics[width=0.8\columnwidth]{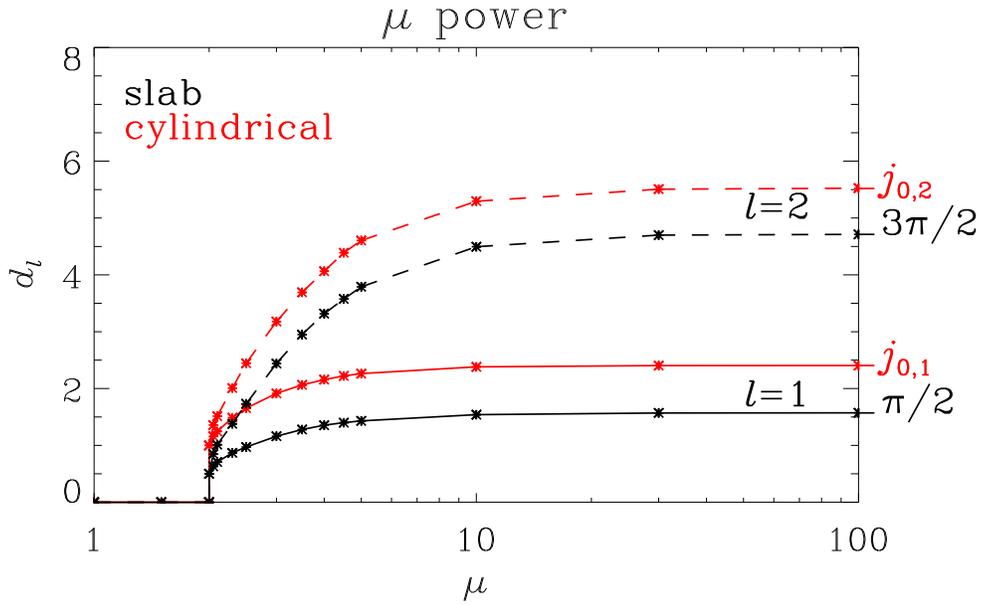}
 \caption{
 Dependence of $d_l$ on $\mu$ for coronal structures with ``$\mu$ power'' profiles.
 Here $d_l$ is related to cutoff axial wavenumbers $\kcl$ by $d_l = (\kcl R)\sqrt{\rhoi/\rhoe-1}$
      and does not depend on the density contrast $\rhoi/\rhoe$.
 Both coronal slabs (the black curves) and tubes (red) are examined.
 The solid and dashed curves represent the transverse fundamental mode (with transverse order $l=1$) 
      and its first harmonic ($l=2$), respectively.
 The horizontal bars present the values of $d_l$ expected for top-hat profiles.
 Note that for ``$\mu$ power'' profiles, $f(x)$ conforms to $1-(x/R)^{\mu_0}$ when $x/R \ll 1$
     and $(x/R)^{-\mu_\infty}$ when $x/R \gg 1$, with $\mu_0 = \mu_\infty = \mu$. 
 See text for details.    
}
 \label{fig_monomu_compare_dl}
\end{figure}

As indicated by Equation~\eqref{eq_kcl_general}, 
    $d_l$ is entirely determined by geometry for a given $f(x)$.
Figure~\ref{fig_monomu_compare_dl} compares the values of $d_l$ in the slab geometry (the black curves)
    with those in the cylindrical one (red)
    for an extensive range of the $\mu$ values.
The solid curves are for the transverse fundamental mode ($l=1$),
    whereas the dashed ones are for the first transverse harmonic ($l=2$).
Given by the horizontal bars are the analytical expectations for $d_l$
    in the top-hat case, namely $\mu=\mu_\infty \rightarrow \infty$.
One sees that $d_l$ is nonzero only when $\mu = \mu_\infty \ge 2$.
Furthermore, while in general $d_l$ is $\mu$ dependent for a given geometry,
    it is not so when $\mu = 2$.
Comparing the black and red curves, one also sees that $d_l$ is always larger
    in the cylindrical geometry at arbitrary $\mu$, indicating that
    {  magnetic slabs with ``$\mu$ power'' profiles can trap sausage waves with longer axial wavelengths.}

\begin{figure}
\centering
\includegraphics[width=0.8\columnwidth]{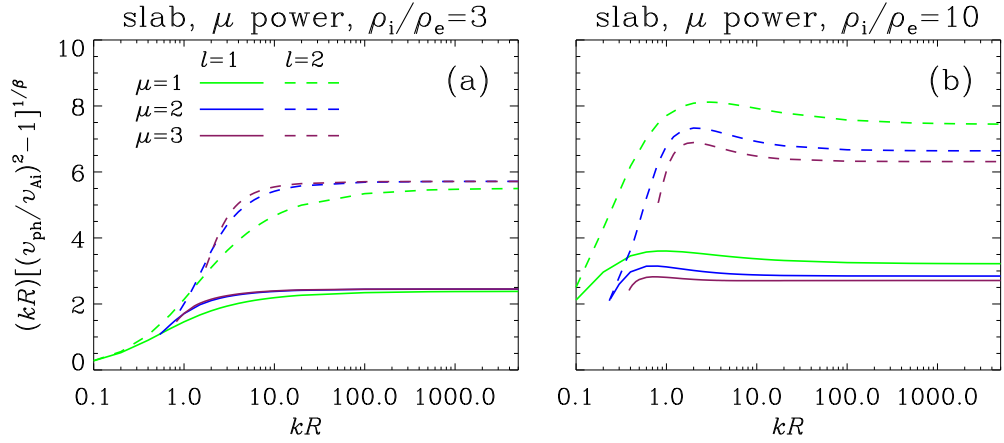}
 \caption{
 Axial phase speed $\vph$ as a function of axial wavenumber $k$
      for trapped sausage waves in coronal slabs with ``$\mu$ power'' profiles.
 The left (right) panel pertains to a density contrast $\rhoi/\rhoe$ of $3$ ($10$).
 Both the transverse fundamental mode (with transverse order $l=1$, the solid curves)
      and its first harmonic ($l=2$, dashed) are examined.
 The results for a number of values of $\mu$ are shown by the curves in different colors.         
 Note that the combination $(kR) [(\vph/\vai)^2-1]^{1/\beta}$ is plotted rather than $\vph$ itself,
      where $\beta = 2\mu/(\mu+2)$.
 Note further that for ``$\mu$ power'' profiles, $f(x)$ conforms to $1-(x/R)^{\mu_0}$ when $x/R \ll 1$
     and $(x/R)^{-\mu_\infty}$ when $x/R \gg 1$, with $\mu_0 = \mu_\infty = \mu$. 
 See text for details.    
}
 \label{fig_monomu_slab_det_cl}
\end{figure}

Now turn to the asymptotic behavior of the axial phase speeds $\vph$ at large axial wavenumbers ($kR$).
Before examining the differences in the two geometries,
    we first employ the slab computations to demonstrate that Equation~\eqref{eq_monomu_vphbigK} to \eqref{eq_def_hl} indeed hold.
To do this, we have evaluated the 
    dependence on $kR$ of $(kR)(\vph^2/\vai^2-1)^{1/\beta}$ for an extensive range of $[l, \mu, \rhoi/\rhoe]$,
    where $\beta$ is given by Equation~\eqref{eq_def_beta} and we take $\mu_0 = \mu$.
A subset of this parameter study is shown in Figure~\ref{fig_monomu_slab_det_cl},
    where we present how $(kR)(\vph^2/\vai^2-1)^{1/\beta}$ depends on $kR$
    for a density contrast $\rhoi/\rhoe=3$ (the left panel) and $\rhoi/\rhoe=10$ (right).
The examined combinations of $[l, \mu$] are represented with different linestyles and colors as labeled.
The point here is that for any given $[l, \mu]$, the curves pertinent to both density contrasts 
    tend to some asymptotic values for sufficiently large $kR$, 
    meaning that Equation~\eqref{eq_monomu_vphbigK} is indeed valid.
In fact, this is how we determine the constant $c_l$.
Comparing the dashed with the solid curves, one sees that $c_l$ is different for different $l$.
Furthermore, $c_l$ also depends on $\mu$ at a given $\rhoi/\rhoe$, even though this dependence is rather weak 
    for a modest density contrast $\rhoi/\rhoe=3$ (see Figure~\ref{fig_monomu_slab_det_cl}a).
When both $l$ and $\mu$ are fixed, $c_l$ is different at different values of $\rhoi/\rhoe$ as can be seen 
    if one compares, say, the asymptotic value that the blue dashed curve attains in Figure~\ref{fig_monomu_slab_det_cl}a
    with that in Figure~\ref{fig_monomu_slab_det_cl}b.

\begin{figure}
\centering
\includegraphics[width=0.8\columnwidth]{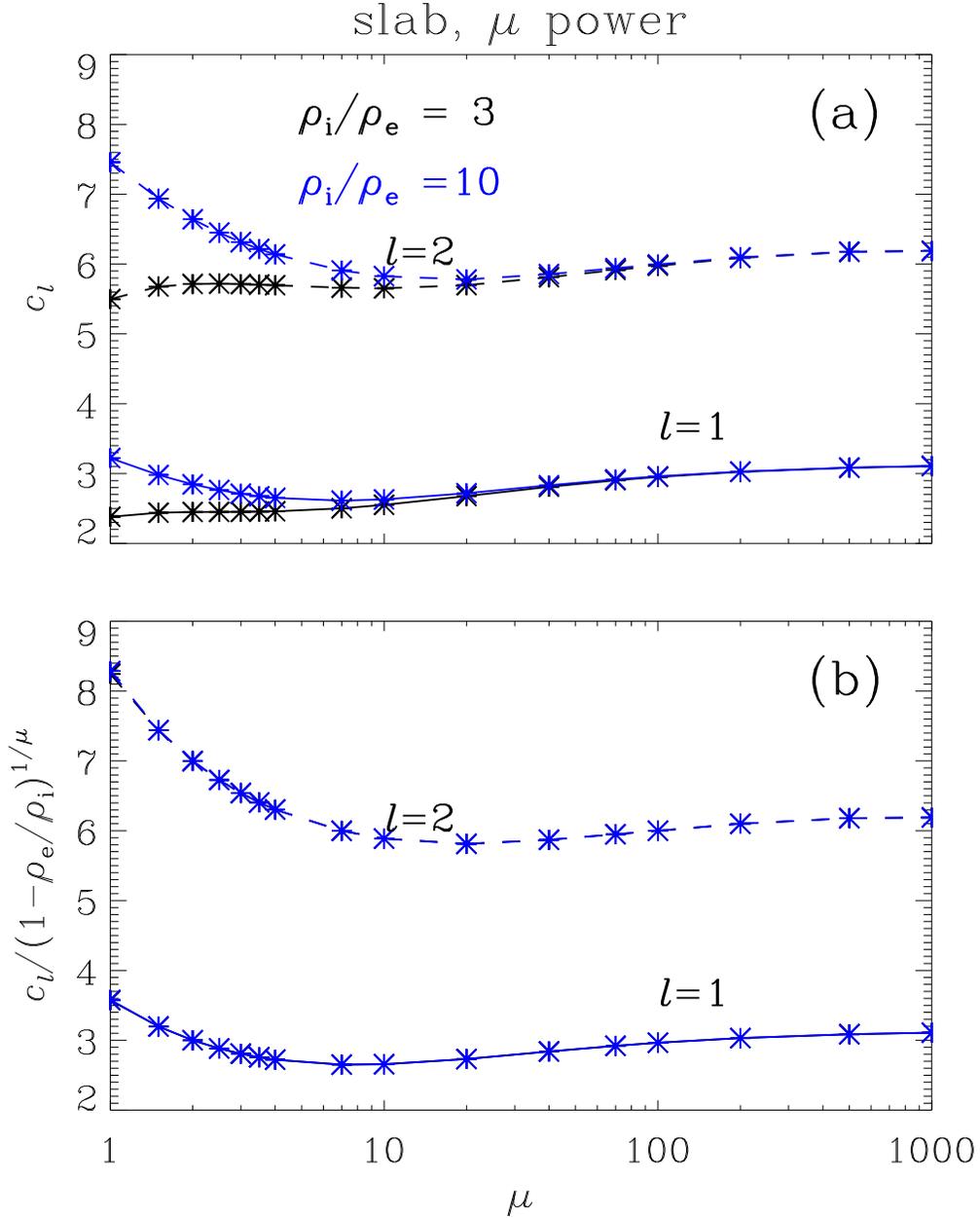}
 \caption{
 Dependence on $\mu$ of $c_l$ (the upper panel) and $c_l/(1-\rhoe/\rhoi)^{1/\mu}$ (lower) for 
      coronal slabs with ``$\mu$ power'' profiles.
 The solid (dashed) curves are for the transverse fundamental mode ($l=1$)
      and its first harmonic ($l=2$), respectively.
 Two density contrasts $\rhoi/\rhoe$ are examined, one being $3$ (the black curves)
      and the other being $10$ (blue).
 Note that for ``$\mu$ power'' profiles, $f(x)$ conforms to $1-(x/R)^{\mu_0}$ when $x/R \ll 1$
     and $(x/R)^{-\mu_\infty}$ when $x/R \gg 1$, with $\mu_0 = \mu_\infty = \mu$. 
 See text for details.    
}
 \label{fig_monomu_slab_det_hl}
\end{figure}

That $c_l$ depends on the combination $[l, \mu, \rhoi/\rhoe]$
    is made more evident
    in Figure~\ref{fig_monomu_slab_det_hl}a, where $c_l$ is shown as a function of $\mu$
    for both $l=1$ (the solid curves) and $l=2$ (dashed)
    and for both $\rhoi/\rhoe=3$ (the black curves) and $\rhoi/\rhoe=10$ (blue).
One sees that $c_l$ increases with $l$ regardless of $\mu$ or $\rhoi/\rhoe$.
Likewise, $c_l$ increases with $\rhoi/\rhoe$ at a given $l$, and this tendency 
    becomes increasingly weak with $\mu$.
This rather complicated dependence on $\rhoi/\rhoe$ makes the comparison of the asymptotic behavior 
    between the two geometries rather cumbersome.
Fortunately, this dependence can be removed if one examines $c_l/(1-\rhoe/\rhoi)^{1/\mu}$.
Shown in Figure~\ref{fig_monomu_slab_det_hl}b, 
    this ratio at a given $l$ becomes solely determined by $\mu$
    as evidenced by the fact that the blue and black curves coincide.
This means that Equation~\eqref{eq_def_hl} holds.
Repeating the same practice involved in Figures~\ref{fig_monomu_slab_det_cl} and \ref{fig_monomu_slab_det_hl} for the cylindrical case,
    we find exactly the same behavior.
    
\begin{figure}
\centering
\includegraphics[width=0.8\columnwidth]{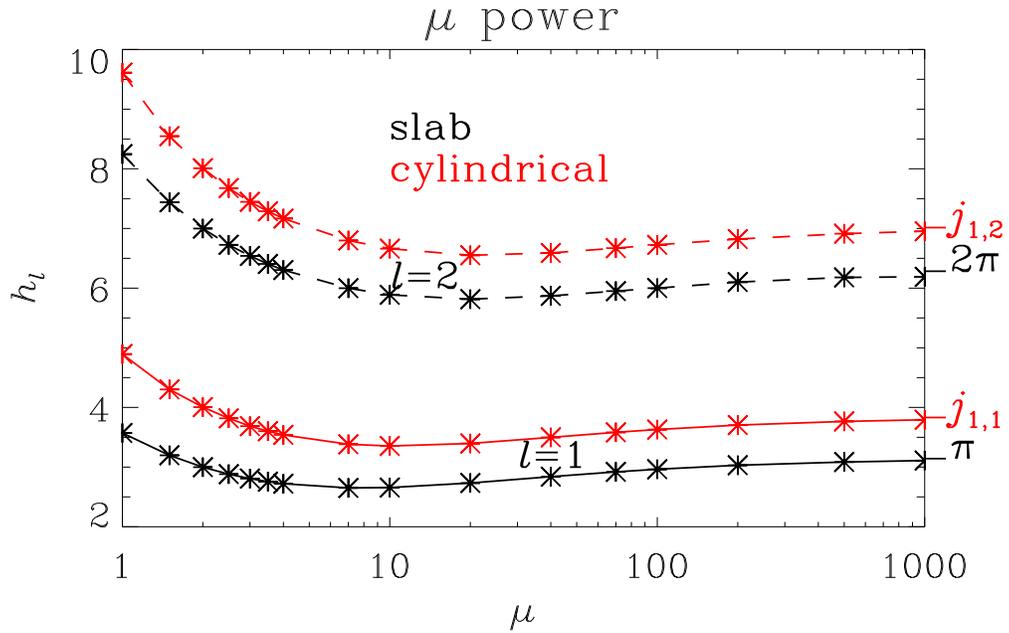}
 \caption{
 Comparison of the dependence of $h_l$ on $\mu$ for trapped sausage waves in 
      coronal slabs (the black curves) and tubes (red). 
 The solid and dashed curves are for the transverse fundamental mode ($l=1$)
      and its first harmonic ($l=2$), respectively.
 The horizontal bars represent the values of $h_l$ expected for top-hat profiles.      
 Note that for ``$\mu$ power'' profiles, $f(x)$ conforms to $1-(x/R)^{\mu_0}$ when $x/R \ll 1$
     and $(x/R)^{-\mu_\infty}$ when $x/R \gg 1$, with $\mu_0 = \mu_\infty = \mu$. 
 See text for details.    
}
 \label{fig_monomu_compare_hl}
\end{figure}

Now we are in a position to compare how the asymptotic behavior of the axial phase speed differs
    in different geometries.
Figure~\ref{fig_monomu_compare_hl}
     shows $h_l = c_l/(1-\rhoe/\rhoi)^{1/\mu}$ as a function of $\mu$ 
     for both the slab (the black curves) and cylindrical (red) geometries
     and for both $l=1$ (the solid curves) and $l=2$ (dashed).
The horizontal bars represent the top-hat
     results pertaining to $\mu \rightarrow \infty$, in which case
     $h_l = l \pi$ ($h_l = j_{1,l}$) for the slab (cylindrical) geometry (see Table~\ref{tbl_3}).
Consider the slab results as given by the black curves.
For ``$\mu$ power'' profiles, the only analytically tractable case happens when $\mu=1$, for which
    Equation~\eqref{eq_monomu1_vphBigK} offers an explicit expression for $h_l$.
Evaluating this $h_l$ yields that $h_1 = 3.53$ and $h_2 = 8.25$, which are in close agreement with the numerical results.
With increasing $\mu$, one sees that $h_l$ decreases first before increasing toward the top-hat results eventually. 
This behavior is also seen in the cylindrical computations shown by the red curves.
It is just that $h_l$ is always larger.

\section{CORONAL SLABS WITH ``OUTER $\mu$'' PROFILES}
\label{sec_outermu}
This section examines coronal slabs with ``outer $\mu$'' profiles as given by Equation~(\ref{eq_profile_outermu}).
For this family of profiles, closed-form DRs can be found for $\mu=1$ and $\mu=2$.
In what follows, we will start with an examination on these two particular choices of $\mu$
   and see what we can expect.

\subsection{Analytical Results for $\mu = 1$ and $\mu = 2$}
\label{sec_sub_EVP_outermu}

Let us start with the case where $\mu = 1$.
The solution to Equation~(\ref{eq_Fourier_xi}) in the outer portion can be shown to have the form
   \footnote{Another independent solution $M_{\nu, 1/2}(X)$ diverges when $x/R$ or equivalently $X$ approaches $\infty$.}
\begin{eqnarray*}
     \tilde{\xi} \propto W_{\nu, 1/2}(X),
\end{eqnarray*}
    where $X= 2 m x$, and $\nu$ is also defined by Equation~(\ref{eq_monomu1_def_nu}).
With $\tilde{\xi}$ expressible in terms of $\sin(n x)$ in the inner portion, the DR reads
\begin{eqnarray}
   nR \displaystyle\cot(nR)
 = mR - \nu -\displaystyle\frac{W_{\nu+1, 1/2}(2 mx)}{W_{\nu, 1/2}(2 mx)}~.
\label{eq_outermu1_DR}
\end{eqnarray}
We have used the fact that
\begin{equation*}
 \displaystyle\frac{\mathd}{\mathd X} W_{\nu, 1/2}(X) =
   \left(\frac{1}{2}-\frac{\nu}{X}\right)W_{\nu, 1/2}(X)
   -\frac{W_{\nu+1, 1/2}(X)}{X}~.
\end{equation*}

No cutoff wavenumbers exist because $f(x)$ at large distances drops less rapidly than $(x/R)^{-2}$.
Some approximate results can be found in the limiting cases where $k\rightarrow 0$ or $k\rightarrow \infty$.
When $k\rightarrow 0$, it turns out that $\nu \rightarrow l$
   for transverse order $l$ with $l = 1, 2, 3, \cdots$.
Actually this is what happens for ``$\mu$ power'' profiles with $\mu=1$, meaning that
   $\vph$ and $\vgr$ can still be approximated by Equations~(\ref{eq_monomu1_vph_smallK})
   and (\ref{eq_monomu1_vgr_smallK}), respectively.
On the other hand, when $kR$ approaches infinity, the RHS of Equation~(\ref{eq_outermu1_DR}) tends to infinity.
As happens in the top-hat case,
    the phase and group speeds for trapped modes of transverse order $l$
    can still be approximated by Equations~(\ref{eq_tophat_vphBigK})
    and (\ref{eq_tophat_vgBigK}), respectively.

Now consider the case where $\mu = 2$.
The solution to Equation~(\ref{eq_Fourier_xi}) in the outer portion can be shown to have the form
   \footnote{Another independent solution $I_{\nu}(X)$ diverges when $x/R$ or equivalently $X$ approaches $\infty$.}
\begin{eqnarray*}
     \tilde{\xi} \propto K_{\nu}(X),
\end{eqnarray*}
    where $X= mx$ and $K_\nu$ is modified Bessel's function of the second kind with
\begin{eqnarray}
\displaystyle
   \nu = \sqrt{\frac{1}{4}-\bar{D}}~.
\end{eqnarray}
With $\tilde{\xi}$ expressible in terms of $\sin(n x)$ in the inner portion, the DR reads
\begin{eqnarray}
   \displaystyle (n R) \cot(nR)
 = \frac{1}{2} - \nu - (m R)\displaystyle\frac{K_{\nu-1}(m R)}{K_{\nu}(m R)}~.
\label{eq_DR_outermu2}
\end{eqnarray}

In this case, the cutoff wavenumbers are still given by Equation~(\ref{eq_monomu2_DR_cutoff}).
One then sees that $\nu \to 0$ at the cutoff because $\bar{D} = 1/4$.
Moving away from this cutoff, $\vph k$ increases with $k$, meaning that $\nu$ becomes purely imaginary
   \citep[see][section 10.45, for a discussion of $K_\nu$ of imaginary order]{NIST:DLMF}.
As is the case for $\mu = 1$, the RHS of Equation~(\ref{eq_DR_outermu2}) grows unbounded
   when $kR$ approaches infinity.
This means that once again $\vph$ and $\vgr$ can be approximated by Equations~(\ref{eq_tophat_vphBigK})
    and (\ref{eq_tophat_vgBigK}), respectively.

\subsection{Comparison with the Cylindrical Results}

Note that for this family of profiles, $\mu_0$ is identically infinite and $\mu_\infty$ is indistinguishable from $\mu$.
In other words, $\mu_0 \equiv \infty$ and $\mu_\infty = \mu$.

\begin{figure}
\centering
\includegraphics[width=0.8\columnwidth]{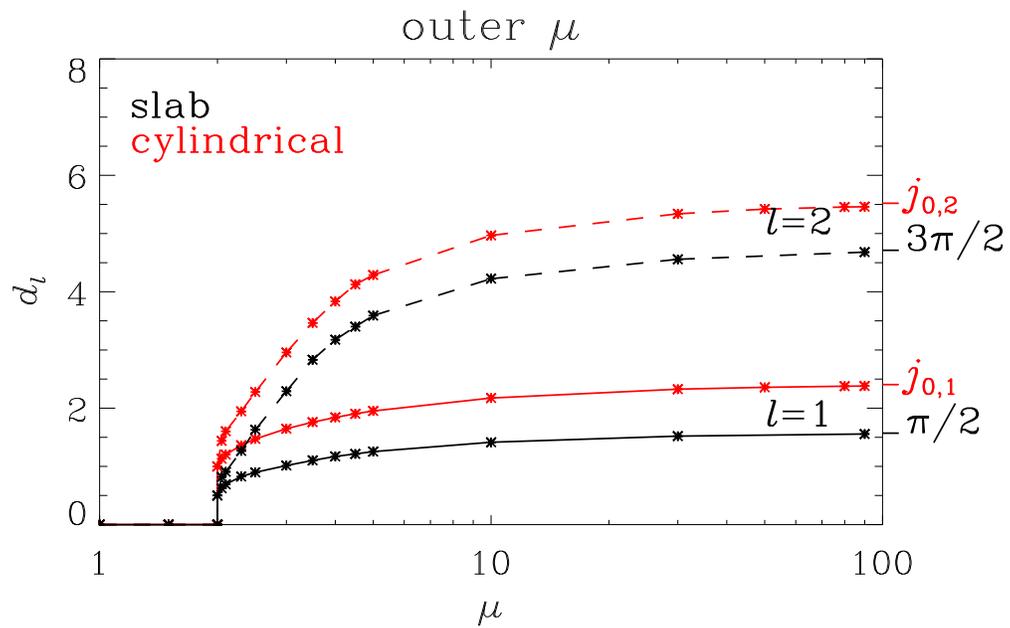}
 \caption{
 Similar to Figure~\ref{fig_monomu_compare_dl} but for coronal slabs with ``outer $\mu$'' profiles.
 Note that for this family of profiles, $f(x)$ conforms to $1-(x/R)^{\mu_0}$ when $x/R \ll 1$
     and $(x/R)^{-\mu_\infty}$ when $x/R \gg 1$, with $\mu_0 = \infty$ and $\mu_\infty = \mu$. 
}
 \label{fig_outermu_compare_dl}
\end{figure}

Figure~\ref{fig_outermu_compare_dl} shows, in a format similar to Figure~\ref{fig_monomu_compare_dl},
   the dependence {  of $d_l$ on $\mu$} for both the slab (the black curves) and cylindrical (red) geometries.
Despite some quantitative difference, the curves look remarkably similar to their counterparts for the ``$\mu$ power'' family of profiles.
In particular, ones sees that $d_l$ is identically zero for $\mu <2$ but increases with $\mu$ before
   leveling off.
Furthermore, the case with $\mu=2$ is special in the sense that $d_l$ does not depend on $l$.
A comparison between the curves in the same linestyle but in different colors shows that 
   $d_l$ is always larger in the cylindrical case.

\begin{figure}
\centering
\includegraphics[width=0.8\columnwidth]{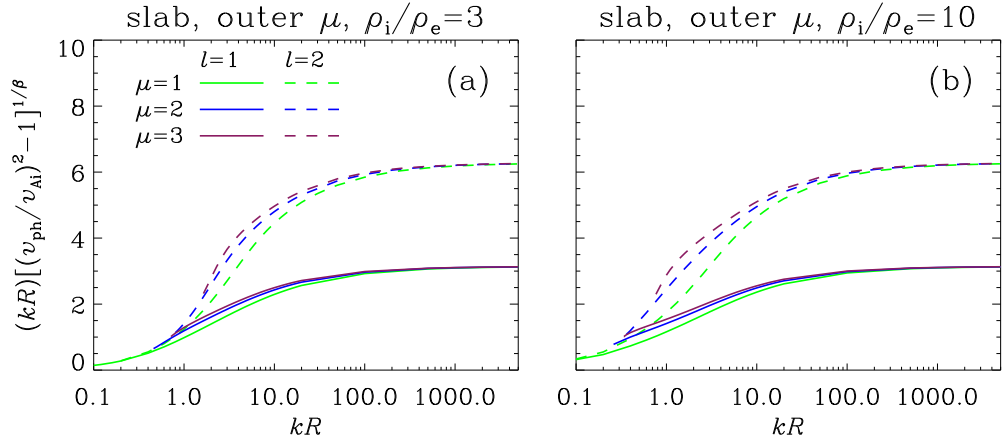}
 \caption{
 Similar to Figure~\ref{fig_monomu_slab_det_cl} but for ``outer $\mu$'' profiles.
 Note that for this family of profiles, $f(x)$ conforms to $1-(x/R)^{\mu_0}$ when $x/R \ll 1$
     and $(x/R)^{-\mu_\infty}$ when $x/R \gg 1$, with $\mu_0 = \infty$ and $\mu_\infty = \mu$. 
 A $\mu_0$ being infinite is employed to evaluate $\beta$, 
     resulting in $\beta=2\mu_0/(\mu_0+2) =2$.
 See text for details.      
}
 \label{fig_outermu_slab_det_cl}
\end{figure}

Let us examine whether Equations~\eqref{eq_monomu_vphbigK} to \eqref{eq_def_hl} hold,     
   still with the slab computations. 
To do this, we can still examine the wavenumber dependence of 
   $(kR) (\vph^2/\vai^2-1)^{1/\beta}$, it is just that now $\beta=2$
   given that $\mu_0 = \infty$ (see Equation~\ref{eq_def_beta}). 
Figure~\ref{fig_outermu_slab_det_cl} shows how $(kR) (\vph^2/\vai^2-1)^{1/\beta}$ varies with $kR$
   for a number of combinations $[l, \mu, \rhoi/\rhoe]$ as labeled.
Similar to the ``$\mu$ power'' profiles (see Figure~\ref{fig_monomu_slab_det_cl}), 
   for sufficiently large $kR$ all the curves
   tend to some constant, which we take as $c_l$.
However, in this case $c_l$ does not depend on the density contrast any more.
In both Figures~\ref{fig_outermu_slab_det_cl}a and \ref{fig_outermu_slab_det_cl}b, we find that 
   $c_l = l\pi$, which is the result expected for top-hat profiles (see Table~\ref{tbl_3}).
In fact, for $\mu=1$ and $\mu=2$, we have analytically shown that $h_l$ attains
   the top-hat values.
That $c_l$ does not depend on $\rhoi/\rhoe$ or $\mu$ 
   lends further support to Conjecture~\ref{conj_vph_bigK}:
   Equation~\eqref{eq_def_hl} indicates that $c_l$ is indistinguishable from $h_l$ 
   and is therefore entirely determined by geometry for a $\mu_0$ being infinite.  
Furthermore, the computations given in Figure~\ref{fig_outermu_slab_det_cl} also shows that 
   the behavior of $f(x)$ away from the slab axis does not play a role in 
   determining $h_l$.
Repeating the computations for the cylindrical geometry yields exactly the same behavior.
It is just that now $c_l$ or equivalently $h_l$ reads $j_{1, l}$ (see Table~\ref{tbl_3}).

\section{CORONAL SLABS WITH ``INNER $\mu$'' PROFILES}
\label{sec_innermu}

This section examines ``inner $\mu$'' profiles as given by Equation~(\ref{eq_profile_innermu}).
In this case, compact closed-form DRs for sausage modes can be found when $\mu=1$ and $2$.

\subsection{Analytical Results for $\mu = 1$ and $\mu = 2$}
\label{sec_sub_EVP_innermu}

Let us start with the case where $\mu=1$.
The solution to Equation~(\ref{eq_Fourier_xi}) in the inner portion is a linear combination of
    Airy's functions $\Ai(X)$ and $\Bi(X)$, where
\begin{eqnarray}
  \displaystyle
  X = \frac{-\bar{n}^2 + \bar{D}^2 (x/R)}{\bar{D}^{2/3}}~.
\end{eqnarray}
Given that $\tilde{\xi}(x = 0) = 0$, the Lagrangian displacement should be of the form
\begin{eqnarray}
  \displaystyle
  \tilde{\xi}(x) \propto \frac{\Ai(X)}{\Ai(X_0)} - \frac{\Bi(X)}{\Bi(X_0)}~,
\end{eqnarray}
   where
\begin{eqnarray}
  \displaystyle
  X_0 = \frac{-\bar{n}^2}{\bar{D}^{2/3}}~
\end{eqnarray}
   is the value of $X$ evaluated at $x=0$.
Note that $\tilde{\xi} \propto {\rm e}^{-mx}$ for $x >R$.
The requirement that ${\mathd \tilde{\xi}}/{\mathd x}$ be continuous at $x=R$ then gives the DR,
\begin{eqnarray}
\displaystyle
\frac{\Bi(X_0) \Ai'(X_1)-\Ai(X_0) \Bi'(X_1)}{\Bi(X_0) \Ai(X_1)-\Ai(X_0) \Bi(X_1)} = -\frac{\bar{m}}{\bar{D}^{1/3}}~,
\label{eq_innermu1_DR}
\end{eqnarray}
    where $\Ai'$ and $\Bi'$ are Airy's prime functions.
Furthermore,
\begin{eqnarray}
  \displaystyle
  X_1 = \frac{\bar{m}^2}{\bar{D}^{2/3}}~
\end{eqnarray}
   is the value of $X$ evaluated at $x=R$.

Cutoff wavenumbers can be derived as follows.
First, one finds that now $X_1 =0$, $\bar{D} = \bar{n}^2$ and
   $X_0 = -\bar{D}^{1/3}$ because $\bar{m} = 0$.
Second, now that the RHS of the DR (\ref{eq_innermu1_DR}) is zero,
   the numerator on the left hand side (LHS) will also be zero.
This means that the cutoff wavenumbers are determined by
\begin{eqnarray}
\displaystyle
\frac{\Ai(X_0)}{\Bi(X_0)} = \frac{\Ai'(0)}{\Bi'(0)} = -\frac{1}{\sqrt{3}} .
\label{eq_innermu1_DR_cutoff}
\end{eqnarray}
{  Defining}
\begin{eqnarray}
\displaystyle
\zeta \equiv \frac{2}{3}\left(-X_0\right)^{3/2},
\end{eqnarray}
    one finds that $\Ai(X_0)/\Bi(X_0)$ is well approximated by $-\cot(\zeta-\pi/4)$ when $|X_0|$ is sufficiently large
    (see AS, pages 448 to 449).
Actually this approximation is accurate to within $2\%$ for $|X_0|$ as small as $2$.
It then follows from Equation~(\ref{eq_innermu1_DR_cutoff}) that $\zeta \approx (l-5/12) \pi$.
With $\zeta = 2/3 \bar{D}^{1/2}$ and $\bar{D} = \kcl^2 R^2 (\rhoi/\rhoe-1)$,
    one eventually finds that
\begin{eqnarray}
 \displaystyle
 \kcl R \approx \frac{(3/2)(l-5/12)\pi}{\sqrt{\rhoi/\rhoe-1}}~,
 \label{eq_innermu1_kc}
\end{eqnarray}
    which is accurate to better than $2\%$.

The asymptotic behavior at large $kR$ can also be analytically established.
For this purpose, we note that both $\bar{D}$ and $\bar{m}^2$ approach infinity.
Furthermore, $\bar{D} \approx \bar{m}^2$, meaning that $X_1 \approx \bar{m}^{2/3}$.
The RHS of Equation~(\ref{eq_innermu1_DR}) is approximately $-\bar{m}^{1/3} \approx -X_1^{1/2}$.
If $\Ai(X_0)$ does not vanish, then one finds that the LHS will be dominated by the terms associated with $\Bi'(X_1)$
    and $\Bi(X_1)$ by using the asymptotic expressions for these two functions at large $X_1$ (see AS, pages 448 to 449).
The consequence is that the LHS is approximately $X_1^{1/2}$, which contradicts the RHS.
This means that the DR at large $kR$ is equivalent to $\Ai(X_0)=0$.
With $X_0 = -\bar{n}^2/\bar{D}^{2/3}$, one then finds that $\vph$ and $\vgr$ can be approximated by
    Equations~(\ref{eq_monomu1_vphBigK}) and (\ref{eq_monomu1_vgBigK}), respectively.
This happens even though these two approximations were derived for the ``$\mu$ power'' profiles with $\mu=1$.

Now consider the case where $\mu=2$.
In this case, the following definitions are necessary,
\begin{eqnarray}
\displaystyle
   && p \equiv \sqrt{\bar{D}} = \frac{\omega R}{\vai} \sqrt{1-\displaystyle\frac{\rhoe}{\rhoi}} ,
   \label{eq_def_innermu2_p}\\
   && \alpha \equiv \frac{1}{4}-\frac{\bar{n}^2}{4p} = \frac{1}{4}-\frac{\left(\omega R/\vai\right)^2-\left(k R\right)^2}{4 p} .
   \label{eq_def_innermu2_alpha}
\end{eqnarray}
The solution to Equation~(\ref{eq_Fourier_xi}) in the inner portion is proportional to
    $X^{1/2} {\rm e}^{-X/2} M(\alpha+1/2, 3/2, X)$ where $X= p (x/R)^2$ and $M(\cdot, \cdot, \cdot)$ is Kummer's M function.
    \footnote{Note that another independent solution ${\rm e}^{-X/2} M(\alpha, 1/2, X)$ is not acceptable
    because it yields a value of unity at $x=0$.}
With $\tilde{\xi}$ expressible in terms of ${\rm e}^{-m x}$ in the outer portion, the DR then reads
\begin{eqnarray}
\displaystyle
   - m R
 = 1 - p + \frac{4 p (\alpha+1/2)}{3} \frac{M(\alpha+3/2, 5/2, p)}{M(\alpha+1/2, 3/2, p)}~.
\label{eq_innermu2_DR}
\end{eqnarray}
In fact, the DR (\ref{eq_innermu2_DR}) has already been given by Equation~(8) in ER88 (also see the references therein).
Its counterpart for cylindrical geometry was given in \citet{2016ApJ...833...51Y}, where we generalized the original treatment
    by \citet{1965PhFl....8..507P} who assumed that $\rhoe=0$.

Now let us offer some new approximate expressions
    for both the cutoff wavenumbers and the asymptotic behavior of
    the phase and group speeds at large wavenumbers.
In fact, both are related to the fact that the DR for any transverse order $l$
    can be approximated by $\alpha \approx 1/2-l$.
Given that $\vph = \vae$ at the cutoff, one sees that
\begin{eqnarray*}
\displaystyle
&& p      = \kcl R \sqrt{\frac{\rhoi}{\rhoe}-1} ,  \\
&& \alpha = \frac{1}{4} -\frac{\kcl^2 R^2({\rhoi}/{\rhoe}-1)}{4p} = \frac{1}{4}-\frac{p}{4} .
\end{eqnarray*}
This means that the cutoff can be approximated by $p\approx 4 l-1$, or equivalently
\begin{equation}
  k_{{\rm c}, l} R \approx \displaystyle\frac{4 l-1}{\sqrt{\rhoi/\rhoe-1}} .
  \label{eq_innermu2_cutoff}
\end{equation}
It turns out this approximation is increasingly accurate with $l$,
   overestimating the exact values by $33\%$, $11.1\%$, and $6.7\%$
   for $l=1$, $2$, and $3$, respectively.

It can be shown that $\alpha$ is almost exactly $1/2-l$ when $kR \rightarrow \infty$.
To demonstrate this, let $A = \alpha+1/2$, and let $E$ denote the last term on the RHS of Equation~(\ref{eq_innermu2_DR}).
Note that now $\bar{m} \rightarrow \infty$ and $p/\bar{m} \rightarrow 1^+$.
For Equation~(\ref{eq_innermu2_DR}) to hold, $E/p$ needs to become zero.
Using the definition of Kummer's M function, $E/p$ evaluates to
\begin{eqnarray*}
\displaystyle
   \frac{E}{p} = \frac{4}{3}A
   \frac{1+\displaystyle\frac{(A+1)}{(5/2)}\frac{p}{1!}+\frac{(A+1)(A+2)}{(5/2)(7/2)}\frac{p^2}{2!}+\cdots}
        {1+\displaystyle\frac{A}{(3/2)}\frac{p}{1!}+\frac{A(A+1)}{(3/2)(5/2)}\frac{p^2}{2!}+\cdots} ~.
\end{eqnarray*}
One finds that $E/p$ is guaranteed to tend to zero when $p\rightarrow \infty$ provided that $A=0, -1, -2, \cdots$,
   which translates into $\alpha = 1/2-l$ with $l=1, 2, 3, \cdots$.
With the definitions of $\alpha$ and $p$, this simple relation can be recast into
   an equation quadratic in $\omega$ whose solution reads
\begin{eqnarray*}
 \displaystyle
 \frac{\omega R}{\vai} = \frac{4l-1}{2} \sqrt{1-\frac{\rhoe}{\rhoi}} + \sqrt{\left(\frac{4l-1}{2}\right)^2\left(1-\frac{\rhoe}{\rhoi}\right)+k^2 R^2} .
\end{eqnarray*}
Consequently,
\begin{eqnarray}
  \displaystyle
  \frac{\vph}{\vai}
  \approx 1+\frac{4l-1}{2}\frac{\sqrt{1-\rhoe/\rhoi}}{k R}
                +\frac{(4 l-1)^2}{8}\frac{(1-\rhoe/\rhoi)}{(k R)^2} ,
  \label{eq_innermu2_vphBigK}
\end{eqnarray}
   and
\begin{eqnarray}
  \displaystyle
  \frac{\vgr}{\vai}
  \approx 1-\frac{(4l-1)^2}{8}\frac{(1-\rhoe/\rhoi)}{(k R)^2} .
  \label{eq_innermu2_vgBigK}
\end{eqnarray}
In other words, one expects to see that $\vgr$ approaches $\vai$
   from below.

\subsection{Comparison with the Cylindrical Results}

Note that for this family of profiles, $\mu_0$ is indistinguishable from $\mu$ whereas $\mu_\infty$ is identically infinite. 
In other words, $\mu_0 =\mu$ and $\mu_\infty \equiv \infty$.

\begin{figure}
\centering
\includegraphics[width=0.8\columnwidth]{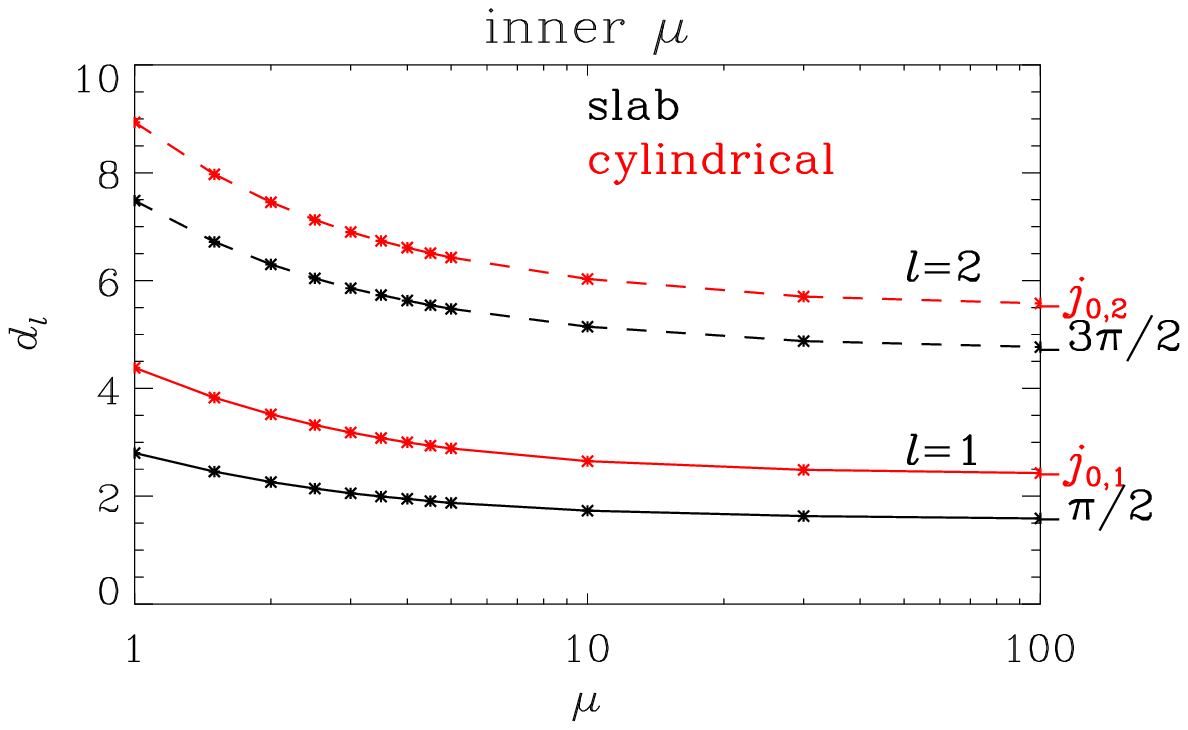}
 \caption{
 Similar to Figure~\ref{fig_monomu_compare_dl} but for ``inner $\mu$'' profiles.
 Note that for this family of profiles, $f(x)$ conforms to $1-(x/R)^{\mu_0}$ when $x/R \ll 1$
     and $(x/R)^{-\mu_\infty}$ when $x/R \gg 1$, with $\mu_0 = \mu$ and $\mu_\infty = \infty$. 
}
 \label{fig_innermu_compare_dl}
\end{figure}

Figure~\ref{fig_innermu_compare_dl} examines, in a format identical to Figure~\ref{fig_monomu_compare_dl},
   how the dependence of $d_l$ on $\mu$ differs in the slab (the black curves) from the cylindrical geometry (red).
One sees that the behavior of $d_l$ is distinct from what happens for the ``$\mu$ power''
   and ``outer $\mu$'' profiles in two aspects.
First, $d_l$ is always nonzero regardless of $\mu$ (or equivalently $\mu_0$).
Second, while $d_l$ tends to increase with $\mu$ for the ``$\mu$ power''
   and ``outer $\mu$'' profiles, it decreases with $\mu$ for the ``inner $\mu$'' profile in both geometries.
This means that the profile details can substantially influence the cutoff wavenumbers, even though 
   whether cutoff wavenumbers exist is entirely determined by the profile steepness far from the structure axis.
Despite these two aspects, one sees once again that 
   $d_l$ is always larger in the cylindrical case.
   
\begin{figure}
\centering
\includegraphics[width=0.8\columnwidth]{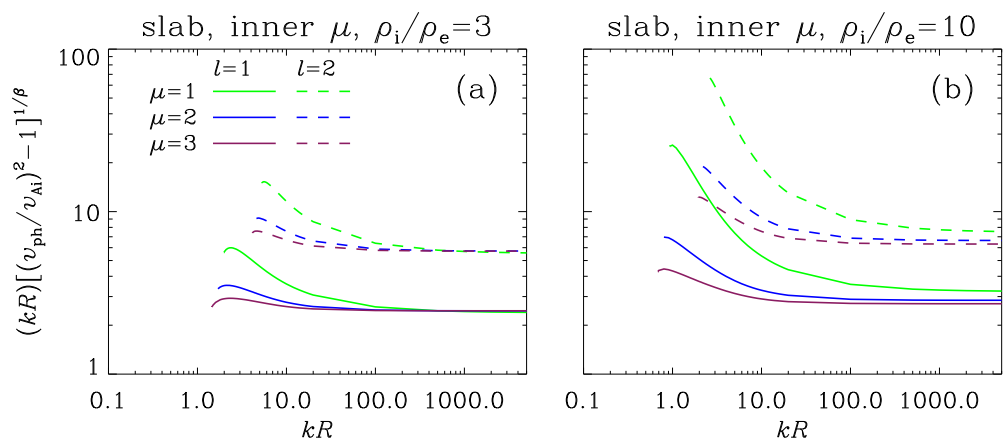}
 \caption{
 Similar to Figure~\ref{fig_monomu_slab_det_cl} but for ``inner $\mu$'' profiles.
 Here $\beta = 2\mu/(\mu+2)$.
 Note that for this family of profiles, $f(x)$ conforms to $1-(x/R)^{\mu_0}$ when $x/R \ll 1$
     and $(x/R)^{-\mu_\infty}$ when $x/R \gg 1$, with $\mu_0 = \mu$ and $\mu_\infty = \infty$. 
}
 \label{fig_innermu_slab_det_cl}
\end{figure}

Do Equations~\eqref{eq_monomu_vphbigK} to \eqref{eq_def_hl} also hold for trapped sausage waves in coronal slabs with the ``inner $\mu$'' profile?
To examine this, Figure~\ref{fig_innermu_slab_det_cl} shows the dependence on $kR$ of $(kR) [(\vph/\vai)^2-1]^{1/\beta}$ in 
    a format identical to Figures~\ref{fig_monomu_slab_det_cl} and \ref{fig_outermu_slab_det_cl}.
Here $\beta$ is given by Equation~\eqref{eq_def_beta} in which we take $\mu_0 = \mu$.
Comparing any curve with its counterpart computed for the ``$\mu$ power'' profile, one sees that 
   both curves attain the same asymptotic value, or equivalently $c_l$, despite that the curves show some
   evident difference when $kR$ is not that large.
In fact, repeating the procedures involved in constructing Figures~\ref{fig_monomu_slab_det_hl}
   and \ref{fig_monomu_compare_hl}, we find exactly the same curves.
The same can be said for the cylindrical case.   
On the one hand, this demonstrates that Equations~\eqref{eq_monomu_vphbigK} to \eqref{eq_def_hl}   
   do hold for this family of profiles, and for coronal slabs and tubes alike.
On the other hand, this lends further support to Conjecture~\ref{conj_vph_bigK} in that 
   the steepness of $f(x)$ at the structure axis is the only factor that determines $h_l$.

\section{SUMMARY AND CONCLUDING REMARKS}
\label{sec_conc}

This study continues our study~\citep[][paper I]{2017ApJ...836....1Y}
    on impulsively generated sausage wave trains 
    in pressureless coronal structures, paying special attention to the effects of the transverse density distribution.
While a cylindrical geometry was examined therein, this study focuses on a slab geometry to capitalize
    on its mathematical simplicity. 
A rather comprehensive survey of representative transverse density profiles
    was conducted, and the temporal and wavelet signatures of impulsively generated wave trains
    were examined in the context of the frequency $\omega$ (or equivalently the axial wavenumber $k$) dependence
    of the axial group speeds ($\vgr$) of trapped modes.
We have also examined the differences that a slab geometry introduces relative to 
    the cylindrical results. 
Our results can be summarized as follows. 

With the much-studied top-hat profile as an example,
    we showed that the temporal evolution and Morlet spectra computed for impulsively generated sausage wave trains
    in coronal slabs are remarkably similar to their cylindrical counterparts.
In fact, they are similar to such an extent that it suffices to refer the readers to the cylindrical results
    as summarized in Table~\ref{tbl_1}.
Common to both geometries, we find that the $\vgr-k$ curves play an essential role in shaping the Morlet spectra.
In particular, the morphology of the Morlet spectra crucially depends on whether the {  group speed} curves
    posses cutoff wavenumbers and whether they are monotonical.
The classical crazy tadpoles are exclusively associated with the {  group speed}
    curves that possess wavenumber cutoffs ($\kcl$)
    and are nonmonotonical.
However, the Morlet spectra in the initial stage in the wave trains are broadened 
    by the appearance of ``fins'' associated with higher-order transverse harmonics 
    when the {  group speed} curves do not possess wavenumber cutoffs. 
Likewise, the Morlet spectra in the late stage tend not to have a broad head
    when the {  group speed} curves
    are monotonical.

We went on to conduct a rather thorough analytical treatment of trapped sausage modes {in coronal slabs}
    for a substantial number of density profiles.
When possible, we provided the analytical expressions for the cutoff wavenumbers (if they exist),
    and the wavenumber dependence of axial phase speeds in the immediate vicinity of these cutoffs
        as well as at sufficiently large wavenumbers. 
These analytical results are summarized in Table~\ref{tbl_4} in a self-contained manner.
\footnote{
    For an $f(x)$ being $1/(1+x/R)^2$, one finds that
        $\vph$ at large $kR$ follows the same form as Equation~(\ref{eq_monomu1_vphBigK}).
    It is just that $c_l$ is twice larger.
    This does not contradict Conjecture~\ref{conj_vph_bigK} because $1/(1+x/R)^2 = 1-2x/R$ for $x/R \ll 1$ to leading order.
}
A thorough numerical analysis of the trapped modes further shows that 
    the density profiles have to fall off no less rapidly than $x^{-2}$ at large distances for cutoff wavenumbers to exist.
On the other hand, the monotonicity of the $\vgr-k$ curves depends critically on 
    the asymptotic behavior of the axial phase speed $\vph$ at sufficiently large $k$,
    which in turn is entirely determined by the profile steepness right at the slab axis.
This asymptotic behavior is quantified by Equations~\eqref{eq_monomu_vphbigK} to \eqref{eq_def_hl},
    which involve only the generic function $f(x)$ that characterizes
    the transition of the density profiles from the internal to ambient values.
Provided that $f(x)$ is approximately $1-(x/R)^{\mu_0}$ when $x \ll R$, 
    our results suggest that $\vgr$ will eventually increase (decrease) with $k$ at large $k$
    when $\mu_0 > 2$ ($\mu_0 < 2$).
The end result is, the $\vgr-k$ curves will be nonmonotonical when $\mu_0 > 2$,
    and tend to be monotical when $\mu_0 < 2$.

The differences between the two geometries are quantified by the behavior of two dimensionless parameters, $d_l$ and $h_l$,
    which do not depend on the density contrast.
Closely related to cutoff wavenumbers, $d_l$ is always larger in the cylindrical case, meaning that 
    {  coronal slabs can trap sausage waves with longer axial wavelengths.}  
On the other hand, $h_l$ is involved in the asymptotic behavior of $\vph$ at large $k$,
    and is found to be also larger in the cylindrical case.
As indicated by Equation~\eqref{eq_monomu_vgrbigK},
    this means that $\vgr$ approaches the internal Alfv\'en speed more rapidly at large $k$
    for sausage waves in coronal tubes than for their counterparts in coronal slabs.

What will be the seismological implications of this rather extensive
    parameter study for both the slab and cylindrical geometries?
Or to be more specific,
    what to make of the Morlet spectra
    that look drastically different from crazy tadpoles?
It is worth emphasizing that high temporal resolution 
    is necessary for these spectra to be observationally found in the first place.
Even for the cases where no cutoff wavenumbers exist and therefore the Morlet spectra extend to substantially longer periods
    than for classical crazy tadpoles, the pertinent periods are at most an order-of-magnitude longer than
    the transverse Alfv\'en time (see Figure~5d in paper I).
For typical coronal structures, this transverse Alfv\'en time evaluates to about one second.
This makes it difficult to discern a proper Morlet spectrum with currently available EUV instruments,
    not to mention the detection of the differences in different types of the spectra.
That said, let us stress that sub-second cadence is readily available in such radio instruments 
    as NoRH (up to 0.1 sec at dual frequencies of 17 and 34 GHz, 
          see \citeauthor{1997LNP...483..183T}~\citeyear{1997LNP...483..183T})
    and the Siberian Solar Radio Telescope 
          (SSRT, 14~ms for one-dimensional imaging observations, 
           see \citeauthor{2003SoPh..216..239G}~\citeyear{2003SoPh..216..239G}).
Furthermore, the solar corona has been imaged in both white light and the coronal green line 
    with a cadence as high as 45~ms with the Solar Eclipse Corona Imaging System
    (SECIS, see \citeauthor{2001MNRAS.326..428W}~\citeyear{2001MNRAS.326..428W}).
It therefore seems possible to dig into the available high-cadence data 
    to look for the Morlet spectra other than crazy tadpoles.
From the results found in the present study, 
    one will be allowed to say that the density profile is likely to be rather gradual in the ambient corona
    if some ``fins'' can be identified.
If, on the other hand, no broad head can be seen, then 
    the density profile close to the structure axis in question
    is likely to be rather steep. 
This latter aspect is perhaps the most important application of the results from our study,
    because it offers a possible means to detect the information on
    the sub-resolution structuring inside coronal structures.

Albeit rather comprehensive, the present study 
    does not exhaust the possible effects on impulsively generated sausage wave trains
    in coronal structures with continuous transverse structuring 
    even when these structures are simply modeled as field-aligned density enhancements.
First, it is evidently impossible to exhaust all possible density descriptions.
Second, in addition to the density profile, the spatial scale of the initial perturbation
    can be equally important in determining 
    the signatures of the wave trains.
As was theoretically shown by \citet{2015ApJ...806...56O}, 
    while the transverse density distribution determines the mode structures,
    the details of the initial perturbations will determine how the energy contained therein
    is apportioned to different modes.
And this energy partition then determines the relative importance of different modes
    in contributing to the temporal evolution of the wave trains.
Pursuing this aspect will provide a more complete picture on impulsively generated wave trains,
    but is left for another study in this series of manuscripts.

\acknowledgments
{  We thank the referee for his/her constructive comments.}
This work is supported by
    the National Natural Science Foundation of China (BL:41474149, 41674172, and 11761141002, HY:41704165,
    SXC:41604145),
    and by the Provincial Natural Science Foundation of Shandong via Grant ZR2016DP03 (HY).

\bibliographystyle{apj}
\bibliography{impls_slab}

\IfFileExists{\jobname.bbl}{} {\typeout{}
\typeout{****************************************************}
\typeout{****************************************************}
\typeout{** Please run "bibtex \jobname" to obtain} \typeout{**
the bibliography and then re-run LaTeX} \typeout{** twice to fix
the references !}
\typeout{****************************************************}
\typeout{****************************************************}
\typeout{}}


\begin{center}
{\bf APPENDIX}
\end{center}

\appendix
\section{CORONAL SLABS WITH EXPONENTIAL PROFILES}
\label{sec_app_EVP_exp}

This section examines coronal slabs with transverse density distributions of the form
\begin{eqnarray}
\displaystyle
 f(x) = \exp\left(-\frac{x}{R}\right)~,
\label{eq_profile_expoential}
\end{eqnarray}
   which has been examined by ER88
   (see \citeauthor{1973ApPhL..23..328C}~\citeyear{1973ApPhL..23..328C} and \citeauthor{1979IJQE...15...14L}~\citeyear{1979IJQE...15...14L}
   for analogous studies in the context of optical fibers).
The solution to Equation~(\ref{eq_Fourier_xi}) has the form
   \footnote{Another independent solution $Y_{2\bar{m}}(X)$ diverges when at $x\rightarrow \infty$ (hence $X \rightarrow 0$).}
\begin{eqnarray}
 \tilde{\xi} \propto J_{2\bar{m}}(X)~,
\end{eqnarray}
   where $X = 2\sqrt{\bar{D}}{\rm e}^{-x/(2R)}$.
The DR for trapped sausage modes is given simply by the requirement that $\tilde{\xi} (x=0) = 0$, resulting in
\begin{eqnarray}
 J_{2\bar{m}} (2\sqrt{\bar{D}}) = 0~.
 \label{eq_exp_DR}
\end{eqnarray}
The cutoff wavenumbers are then given by
\begin{eqnarray}
\displaystyle
  \kcl = \frac{j_{0, l}/2}{\sqrt{\rhoi/\rhoe - 1}}~,
 \label{eq_exp_kc}
\end{eqnarray}
   which follows from $\bar{m} = 0$ and $\bar{D} = k^2 R^2 (\rhoi/\rhoe-1)$.
The expressions for both the DR (\ref{eq_exp_DR}) and cutoff wavenumbers (\ref{eq_exp_kc}) have already been given by ER88.
We will further examine the dispersive properties of sausage modes both in the immediate vicinity of cutoff wavenumbers and
   at large axial wavenumbers.

To show what happens when $k$ exceeds $\kcl$ by only a small amount, we start by noting that Equations~(\ref{eq_def_deltak_deltaomega})
   to (\ref{eq_def_deltan_m_2_deltakomg}) remain valid.
Let $\Dbarcl$ denote the value that $\bar{D}$ attains at $\kcl$, which evaluates to $\nbarcl^2$.
It is straightforward to show that $\bar{D}\approx \Dbarcl (1+2\delta_\omega)$.
Now expanding the DR (\ref{eq_exp_DR}) by seeing the order and argument as being independent
   from each other, one finds that
\begin{eqnarray}
\displaystyle
   \left(2\bar{m}\right)\left[\frac{\partial}{\partial \nu}J_\nu\left(2\sqrt{\bar{D}_{{\rm c}, l}}\right)\right]_{\nu=2\bar{m}\rightarrow 0}
+  2\left(\sqrt{\bar{D}}-\sqrt{\Dbarcl}\right)\left[\frac{\partial}{\partial Z}J_0\left(Z\right)\right]_{Z=2\sqrt{\Dbarcl}}
  = 0.
\end{eqnarray}
Note that $\partial J_\nu/\partial \nu|_{\nu\rightarrow 0} = (\pi/2)Y_0$
   and $\partial J_0(Z)/\partial Z = -J_1(Z)$ (see AS, chapter 9).
Some algebra shows that $\delta_\omega$ is related to $\delta_k$ by
   an equation in the same form as Equation~(\ref{eq_link_deltaomg_deltak}) with $\eta$
   now given by $\eta = (2/\pi)[J_1(j_{0,l})/Y_0(j_{0,l})]$.
Consequently, the approximate expressions for the phase and group speeds are still given by
   Equations~(\ref{eq_tophat_vph_kcvic}) and (\ref{eq_tophat_vgr_kcvicinity}), respectively.

When $kR \rightarrow \infty$, it is easy to see that
    $\chi \equiv \sqrt{\bar{D}}/\bar{m} \rightarrow 1^+$ despite that
    both $\bar{D}$ and $\bar{m}$ approach infinity.
Now let $\zeta$ denote the solution to
\begin{eqnarray}
 \displaystyle
 \frac{2}{3}\left(-\zeta\right)^{3/2} = \sqrt{\chi^2-1} -{\rm arcsec}\chi~.
 \label{eq_exp_def_zeta}
\end{eqnarray}
The uniform asymptotic expansion for $\bar{m} \rightarrow \infty$ through real values
    indicates that $J_{2\bar{m}}(2\bar{m} \chi)$ is dominated by a term associated with $\Ai[(2\bar{m})^{2/3}\zeta]$
    \citep[see][Equation 10.20.4]{NIST:DLMF}.
This means that asymptotically $\zeta = a_l (2\bar{m})^{-2/3}$ with $a_l$ being the zeros of Airy's function $\Ai$.
On the other hand, letting $\chi^2 = 1/(1-\delta)$ with $0< \delta \ll 1$, one finds from Equation~(\ref{eq_exp_def_zeta}) that
    $\delta = -2^{2/3}\zeta$.
With the definitions given by Equation~(\ref{eq_def_nmD_nondimen}), one then finds that
    $\vph^2/\vai^2 = 1+(1-\rhoe/\rhoi)\delta$.
Putting all these results together, we finally find that the approximate expressions for $\vph$ and $\vgr$ at large $kR$
    agree exactly with Equations~(\ref{eq_monomu1_vphBigK}) and (\ref{eq_monomu1_vgBigK}), respectively.

\section{CORONAL SLABS WITH SYMMETRIC EPSTEIN PROFILES}
\label{sec_app_EVP_eps}

This section examines coronal slabs with transverse density distributions of the form
\begin{eqnarray}
\displaystyle
 f(x) = {\rm sech}^2\left(\frac{x}{R}\right)~,
\label{eq_profile_epstein}
\end{eqnarray}
   which is the symmetric Epstein profile.
While the DR for trapped sausage modes in this case can be found in e.g., \citet{2003A&A...409..325C}, \citet{2011A&A...526A..75M},
    and \citet{2014SoPh..289.1663C},
    we think a detailed derivation will be informative.
Furthermore, we will derive the expressions for both the cutoff wavenumbers and the asymptotic behavior
    of the phase and group speeds for arbitrary transverse order $l$,
    which are not available to our knowledge.

An equation identical in form to Equation~(\ref{eq_Fourier_xi}) was originally treated
    in \citet[][page 73]{1965qume.book.....L} in the context of quantum mechanics
    (see \citeauthor{1979IJQE...15...14L}~\citeyear{1979IJQE...15...14L} for an analogous study treating optical fibers).
With the definitions
\begin{eqnarray}
\displaystyle
  X = \tanh\left(x/R\right)~, \label{eq_eps_def_X}\\
  \nu = \sqrt{\bar{D}+\frac{1}{4}}-\frac{1}{2}~, \label{eq_eps_def_nu}
\end{eqnarray}
  Equation~(\ref{eq_Fourier_xi}) becomes
\begin{eqnarray}
\displaystyle
    \frac{\mathd}{\mathd X}\left[\left(1-X^2\right)\frac{\mathd \tilde{\xi}}{\mathd X}\right]
   +\left[\nu\left(\nu+1\right)-\frac{\bar{m}^2}{1-X^2}\right]\tilde{\xi} = 0~,
\label{eq_eps_xi}
\end{eqnarray}
   which is the associated Legendre equation.
While \citet{1965qume.book.....L} went on to transform Equation~(\ref{eq_eps_xi}) into the hypergeometric equation,
   we find that it is slightly more convenient to stick to this present form for examining
   sausage waves.
The solution to Equation~(\ref{eq_eps_xi}) can be expressed as
\begin{eqnarray*}
  \tilde{\xi} \propto C_1 P_{\nu}^{\bar{m}}(X) + C_2 Q_{\nu}^{\bar{m}}(X)~,
\end{eqnarray*}
   where $P_\nu^{\bar{m}}$ and $Q_\nu^{\bar{m}}$ are the associated Legendre functions (AS, Chapter~8).
For $\tilde{\xi}$ to vanish at $x\rightarrow \infty$ (and hence $X=1^-$),
   the two constants $C_1$ and $C_2$ need to be related by $C_1 = -(C_2/2)\pi\cot(\bar{m}\pi)$.
Evaluating $\tilde{\xi}$ at $x= 0$ (and hence $X=0$) yields that
\begin{eqnarray}
\displaystyle
   \tilde{\xi} (X=0) = -C_2 \pi^{3/2}\frac{{\rm csc}(\bar{m}\pi)\Gamma(\bar{m}+\nu+1)}
   {\Gamma[(\bar{m}-\nu+1)/2]\Gamma(\nu-\bar{m}+1)\Gamma[(\bar{m}+\nu)/2+1]}~,
\end{eqnarray}
   where $\Gamma$ is the Gamma function.
Given that $\tilde{\xi}(X=0) = 0$ for sausage modes,
   $\Gamma[(\bar{m}-\nu+1)/2]$ needs to diverge, meaning that
   $(\bar{m}-\nu+1)/2 = 0, -1, -2, \cdots$.
In other words, the desired DR reads
\begin{eqnarray}
\displaystyle
\sqrt{\bar{D}+\frac{1}{4}} = \bar{m}+\left(2l-\frac{1}{2}\right)~,
\label{eq_eps_DRprelim}
\end{eqnarray}
   with $l=1, 2, 3, \cdots$ being the transverse order.
We note by passing that with $C_1$ and $C_2$ related in the above-mentioned manner, $\tilde{\xi}$ can be transformed into
\begin{eqnarray}
\displaystyle
 \tilde{\xi} \propto \left(1-X^2\right)^{\bar{m}/2} F\left(\bar{m}+\nu+1, \bar{m}-\nu; \bar{m}+1; \frac{1-X}{2}\right)~,
 \label{eq_eps_xi_hyper}
\end{eqnarray}
   where $F(a, b; c; \cdot)$ denotes the hypergeometric function $\,_2F_1$.
Equation~(\ref{eq_eps_xi_hyper}) is identical to Equation~(10) in \citet{2011A&A...526A..75M}.

Solving Equation~(\ref{eq_eps_DRprelim}) for $\omega^2$ can make
     the dependence of $\omega$ on $k$  more apparent.
To this end, we first take the squares of both sides, resulting in
\begin{eqnarray}
\displaystyle
  \bar{n}^2-2l(2l-1) = (4l-1)\bar{m}~.
  \label{eq_eps_n2_vs_m}
\end{eqnarray}
It then follows that
\begin{eqnarray}
\displaystyle
&& \frac{\omega^2 R^2}{\vai^2}  \nonumber \\
&=& \left(kR\right)^2
     +\frac{4l(2l-1)-(4l-1)^2\rhoe/\rhoi}{2} \nonumber  \\
&+&
      \frac{(4l-1)\sqrt{4(1-\rhoe/\rhoi)(kR)^2+[(4l-1)^2(\rhoe/\rhoi)^2-8l(2l-1)\rhoe/\rhoi]}}{2}~.
\label{eq_eps_DR}
\end{eqnarray}
Furthermore, evaluating Equation~(\ref{eq_eps_DRprelim}) at the cutoff ($\vph = \vae$ and hence $\bar{m} = 0$) yields a compact expression for
    the cutoff wavenumbers
\begin{eqnarray}
\displaystyle
   \kcl R = \frac{\sqrt{2l(2l-1)}}{\sqrt{\rhoi/\rhoe-1}}~.
\label{eq_eps_kc}
\end{eqnarray}
We note that the term in the square root on the RHS of Equation~(\ref{eq_eps_DR}) can be positive
    even when $k$ is smaller than $\kcl$.
However, the solution in the portion $k < \kcl$ is spurious: plugging it in the LHS of Equation~(\ref{eq_eps_n2_vs_m})
    yields a negative value.
This is not acceptable because with $\bar{m}$ (and $m$) negative, Equation~(\ref{eq_eps_xi_hyper}) indicates that
    the Lagrangian displacement $\tilde{\xi}$ becomes unbounded with distance, given that
    $X\rightarrow 1^-$ and $F[\cdot, \cdot; \cdot; (1-X)/2] \rightarrow 1$ when $x\rightarrow \infty$.

Approximate expressions for both $\vph$ and $\vgr$ can be found when $k$ is larger than $\kcl$ only marginally.
The derivation is simpler than for top-hat and exponential profiles because we can now directly plug
    into Equation~(\ref{eq_eps_n2_vs_m}) the definitions in Equations~(\ref{eq_def_deltak_deltaomega})
    to (\ref{eq_def_deltan_m_2_deltakomg}).
One finds that $\delta_\omega$ is related to $\delta_k$ by
   an equation in the same form as Equation~(\ref{eq_link_deltaomg_deltak}) with $\eta$
   now given by $\eta = 2\sqrt{2l(2l-1)}/(4l-1)$.
The approximate behavior of the phase and group speeds can still be described by
   Equations~(\ref{eq_tophat_vph_kcvic}) and (\ref{eq_tophat_vgr_kcvicinity}), respectively.

We now look for the asymptotic behavior of the axial phase and group speeds at large axial wavenumbers.
Starting with Equation~(\ref{eq_eps_DR}), we find that when $kR \rightarrow \infty$,
\begin{eqnarray}
\displaystyle
\frac{\vph}{\vai} &\approx& 1+\frac{(2l-1/2)\sqrt{1-\rhoe/\rhoi}}{kR}
   -\frac{(2l-1/2)^2(1+\rhoe/\rhoi)/2-l(2l-1)}{k^2 R^2}~, \label{eq_eps_vpBigK} \\
\frac{\vgr}{\vai} &\approx& 1+\frac{(2l-1/2)^2(1+\rhoe/\rhoi)/2-l(2l-1)}{k^2 R^2}~. \label{eq_eps_vgBigK}
\end{eqnarray}
One sees that Equations~(\ref{eq_eps_vpBigK}) and (\ref{eq_eps_vgBigK})
    agree with Equations~(\ref{eq_innermu2_vphBigK}) and (\ref{eq_innermu2_vgBigK}),
    if the terms of order $1/(kR)^2$ are neglected.
However, it can be readily shown that the numerator in the second term on the RHS of Equation~(\ref{eq_eps_vgBigK})
    is positive for arbitrary $\rhoe/\rhoi$.
This means that eventually $\vgr$ approaches $\vai$ from above,
    which is opposite to what happens in the inner $\mu$ case with $\mu=2$.

\section{CORONAL SLABS WITH $f(x) = 1/(1+x/R)^2$}
\label{sec_app_EVP_invSqr}

This section examines coronal slabs with transverse density distributions of the form
\begin{eqnarray}
\displaystyle
 f(x) = \frac{1}{(1+x/R)^2}~,
\label{eq_profile_LNmu2}
\end{eqnarray}
   which has already been examined by LN15
   (see \citeauthor{1979IJQE...15...14L}~\citeyear{1979IJQE...15...14L} for an analogous study treating optical fibers).
The solution to Equation~(\ref{eq_Fourier_xi}) is in the form
\begin{eqnarray}
\displaystyle
   \tilde{\xi} \propto X^{1/2} K_\nu(X)~,
\end{eqnarray}
   where
\begin{eqnarray}
\displaystyle
   X = \bar{m} (1+x/R)~,~~~
   \nu^2 = \frac{1}{4}-\bar{D}~.
   \label{eq_LNmu2_def_nu}
\end{eqnarray}
The DR for trapped sausage modes simply follows from the requirement that $\tilde{\xi}(x=0) = 0$, resulting in
\begin{eqnarray}
\displaystyle
  K_\nu (\bar{m}) = 0~.
\label{eq_LNmu2_DR}
\end{eqnarray}
Note that at large distances this profile reads $f(x) \approx R^2/x^2$, which is identical to our ``$\mu$ power'' case with $\mu=2$.
Similar to the discussions therein, the cutoff wavenumbers should be still given by Equation~(\ref{eq_monomu2_DR_cutoff}), and
   $\nu$ is either zero or purely imaginary ($\nu^2 \le 0$).

The results up to this point have already been found by LN15.
Therefore let us now offer some new analytical results by examining what happens when $kR \rightarrow \infty$.
Note that in this case both $\bar{m}$ and $|\nu|$ approach infinity, but $\chi \equiv \bar{m}/|\nu| \rightarrow 1^-$.
Using the asymptotic expansion for the modified Bessel function of purely imaginary order
    \citep[see][page 425, Ex.~10.6]{1997asymp.book.....O},
    we find that $K_{i|\nu|}(|\nu|\chi)$ is dominated by a term proportional to $\Ai(-|\nu|^{2/3} \zeta)$,
    where $\zeta$ is the solution to
\begin{eqnarray}
\displaystyle
 \frac{2}{3}\zeta^{3/2} = \ln \frac{1+(1-\chi^2){1/2}}{\chi} - (1-\chi^2)^{1/2}~.
 \label{eq_LNmu2_zeta}
\end{eqnarray}
This means that asymptotically $\zeta \approx (-a_l) |\nu|^{-2/3}$.
Now let $\chi^2 = 1-\delta$ with $0<\delta\ll 1$.
Taylor expanding Equation~(\ref{eq_LNmu2_zeta}) to terms of order $\delta^{3/2}$, one finds that
   $\delta = 2^{2/3}\zeta$.
In addition, it is simple to show that $\vph^2/\vai^2 = 1+\delta(1-\rhoe/\rhoi)$, meaning that
\begin{eqnarray}
\displaystyle
 \frac{\vph^2}{\vai^2}
 && \approx 1+ (-a_l)2^{2/3}(1-\rhoe/\rhoi)(kR\sqrt{1-\rhoe/\rhoi})^{-2/3} \nonumber \\
 && \approx 1+\left[\frac{3\left(4l-1\right)\pi(1-\rhoe/\rhoi)}{4kR}\right]^{2/3}~,
 \label{eq_LNmu2_vpBigK}
\end{eqnarray}
   where the second approximation follows from the accurate approximation to $a_l$,
   the $l$-th zero of Airy's function $\Ai$ (see Equation~\ref{eq_def_AiZero}).
Consequently, the axial group speed at large $kR$ is given by
\begin{eqnarray}
    \displaystyle
    \frac{\vgr^2}{\vai^2}
    \approx 1+\frac{1}{3}\left[\frac{3\left(4l-1\right)\pi(1-\rhoe/\rhoi)}{4kR}\right]^{2/3}~.
    \label{eq_LNmu2_vgBigK}
\end{eqnarray}
This means that both $\vph$ and $\vgr$ eventually approach $\vai$ from above when $kR$ increases.

\clearpage 
\begin{deluxetable}{c|c|c|c|c|c}
\tablewidth{\textwidth}
\tabletypesize{\scriptsize}
\tablecaption{
Impulsively generated sausage wave trains in coronal slabs\tablenotemark{a}
\label{tbl_1}
}
\startdata
\hline
{profile} & {$f(x)$ at $x/R \ll 1$}	& {$f(x)$ at $x/R \gg 1$}	& {$[\mu_0, \mu_\infty]$} 	& {Group Speed Curves} 		& {Morlet Spectra} \\
\hline  
$\mu$ power
	  & $\approx 1-(x/R)^\mu$	& $\approx (x/R)^{-\mu}$	&  $[\mu, \mu]$ 		& See Fig.~4 in paper I		& See Figs.~5 \& 6 in paper I	\\
\hline  
outer $\mu$
	  & $=1$			& $=(x/R)^{-\mu}$		&  $[\infty, \mu]$ 		& See Fig.~7 in paper I		& See Figs.~8 \& 9 in paper I	\\
\hline  
inner $\mu$	
	  & $=1-(x/R)^\mu$		& $=0$				&  $[\mu, \infty]$ 		& See Figs.~10 \& 11 in paper I	& See Figs.~12 \& 13 in paper I	\\
\enddata
\tablenotetext{a}{\tiny 
As is the case in cylindrical geometry, the behavior of impulsively generated sausage wave trains in 
    coronal slabs depends crucially on the behavior of the group speed curves pertinent to trapped modes.
These group speed curves are qualitatively similar to the cylindrical case, and 
    we refer the readers to their cylindrical counterparts given in paper I.
The same is true for the temporal evolution and Morlet spectra computed for the density perturbations sampled at a distance
    far from the impulsive source.
}

\end{deluxetable}

\clearpage
\begin{deluxetable}{c|p{3.5cm}|p{3.5cm}|p{3.5cm}}
\tablewidth{\textwidth}
\tabletypesize{\footnotesize}
\tablecaption{
Comparison of $d_l$ between the slab and cylindrical geometries\tablenotemark{a} 
\label{tbl_2}
}
\startdata
\hline
{$f(x)$} 						&  
{$l$} 							& 
{Slab} 							& 
{Cylindrical}  						
\\
\hline  
\multirow{2}{*}{top-hat\tablenotemark{b}} 
				& $l=1$		& $\pi/2 =1.57$    & $j_{0,1}=2.41$ \\	
				& $l=2$		& $3\pi/2=4.71$    & $j_{0,2}=5.52$  \\	
				& $l=3$		& $5\pi/2=7.85$    & $j_{0,3}=8.65$  \\	     
\hline  
{$1/(1+x/R)^2$} 	        & {arbitrary $l$} & {$1/2$}        & {$1$} \\	
\hline  
\multirow{2}{*}{$\exp(-x/R)$} 	& $l=1$		& $j_{0,1}/2=1.20$ & $1.95$ \\	
				& $l=2$		& $j_{0,2}/2=2.76$ & $3.44$  \\	
(see note c)				& $l=3$		& $j_{0,3}/2=4.33$ & $4.97$  \\	     
\hline  
\multirow{2}{*}{${\rm sech}^2(x/R)$}
				& $l=1$		& $1.41$ 	   & $2.26$ \\	
				& $l=2$		& $3.46$ 	   & $4.23$  \\	
(see note d)				& $l=3$		& $5.48$ 	   & $6.21$  \\	     
\enddata
\tablenotetext{a}{\tiny The profiles in this table are all analytically tractable in the slab geometry. 
    In the cylindrical one, however, $d_l$ can be found analytically only for the first two profiles.
    For the rest of profiles, $d_l$ is found numerically via BVPSuite.}
\tablenotetext{b}{\tiny In the slab geometry, $d_l = (l-1/2)\pi$. In the cylindrical one,  $d_l = j_{0, l}$.}
\tablenotetext{c}{\tiny In the slab geometry, $d_l = j_{0, l}/2$.}
\tablenotetext{d}{\tiny In the slab geometry, $d_l = \sqrt{2l(2l-1)}$).}

\end{deluxetable}

\clearpage
\begin{deluxetable}{c|p{3.5cm}|p{2.5cm}|p{2.5cm}|p{2.5cm}}
\tablewidth{\textwidth}
\tabletypesize{\footnotesize}
\tablecaption{
Comparison of $h_l$ between the slab and cylindrical geometries\tablenotemark{a} 
\label{tbl_3}
}
\startdata
\hline
{$f(x)$} 						& 
{$f(x)$ at small $x/R$} 				& 
{$l$}							& 
{Slab} 							& 
{Cylindrical}  						
\\
\hline  
{top-hat\tablenotemark{b}} 
				& $1-(x/R)^{\infty}$	& $l=1$		& $\pi =3.14$ 	& $j_{1,1}=3.83$ \\	
				& 			& $l=2$		& $2\pi=6.28$   & $j_{1,2}=7.02$  \\	
\hline  
{$\exp(-x/R)$}			& {$\approx 1-(x/R)$}    & $l=1$	& $3.53$ & $4.89$ \\	
(see note c)			&			 & $l=2$	& $8.25$ & $9.62$  \\	
\hline  
{${\rm sech}^2(x/R)$}		& {$\approx 1-(x/R)^2$}  & $l=1$	& $3$   & $4$ \\	
(see note d)			&			 & $l=2$	& $7$   & $8$  \\	
\enddata
\tablenotetext{a}{\tiny 
    The profiles in this table are all
        analytically tractable in the slab geometry. 
    In the cylindrical one, however, $h_l$ can be found analytically only for the top-hat profile.
    For the rest of profiles, $h_l$ is found numerically via BVPSuite.
    }
\tablenotetext{b}{\tiny In the slab geometry, $h_l = l\pi$. In the cylindrical one, $h_l = j_{1, l}$.}
\tablenotetext{c}{\tiny In the slab geometry, $h_l = 3(4l-1)\pi/8$.}
\tablenotetext{d}{\tiny In the slab geometry, $h_l = 4l-1$. In the cylindrical one, while $h_l$ cannot be analytically found, 
    the numerically derived values are in exact agreement with what we found with inner $\mu$ profiles with $\mu=2$.
    Equation~(30) in paper I indicates that $h_l = 4l$.}
\end{deluxetable}

\clearpage

\begin{deluxetable}{c|l|l|c|c|c}
\tablewidth{\textheight}
\tabletypesize{\scriptsize}
\rotate
\tablecaption{
Summary of analytical behavior of trapped sausage modes in coronal slabs \tablenotemark{a}
\label{tbl_4}
}
\startdata
\hline
\multirow{2}{*}{profile} 			&  
\multirow{2}{*}{$f(x)$} 				& 
\multirow{2}{*}{Dispersion Relation} 		& 
$d_l$ in cutoff  					& 
($c_l, \beta$) in $\vph$ at 				& 
\multirow{2}{*}{References \tablenotemark{d}} 		\\
						& 
						& 
						& 
 wavenumbers \tablenotemark{b}					& 
 large wavenumbers \tablenotemark{c} 						& 
 \\
\hline  
    \multirow{2}{*}{top-hat} 						& 
                    $1$ if $x \le R$							& 
    \multirow{2}{*}{$\bar{n}\cot\bar{n} = -\bar{m}$}   			& 
    \multirow{2}{*}{$(l-1/2)\pi$}					& 
    {$c_l = l\pi$} 					& 
    \multirow{2}{*}{ER82, also Sect.~\ref{sec_sub_EVP_tophat}} \\
			&   
    $0$ if $x > R$	&   
			&   
			&   
	$\beta = 2$	&   
			\\
\hline   
    {$\mu$ power} 						& 
    \multirow{3}{*}{$\displaystyle\frac{1}{1+x/R}$}							& 
    \multirow{2}{*}{$W_{\nu,1/2}(2\bar{m})=0$}   			& 
    \multirow{3}{*}{no cutoff \tablenotemark{e}}					& 
    \multirow{2}{*}{$\displaystyle c_l=\frac{3(4l-1)\pi(1-\rhoe/\rhoi)}{8}$} 					& 
    \multirow{2}{*}{LN15, also Sect.~\ref{sec_sub_EVP_monomu}} \\
    ($\mu=1$)			&   
			&   
	&   
			&   
			&   
			\\
    			&   
			&   
        where $\nu=\bar{D}/(2\bar{m})$		&   
			&   
	$\beta = 2/3$		&   
			\\
\hline   
    {outer $\mu$} 						& 
    $1$ if $x\le R$							& 
    \multirow{2}{*}{$\displaystyle\bar{n}\cot\bar{n}=\bar{m}-\nu-\frac{W_{\nu+1,1/2}(2\bar{m})}{W_{\nu,1/2}(2\bar{m})}$}   			& 
    \multirow{3}{*}{no cutoff \tablenotemark{e}}					& 
    \multirow{2}{*}{$c_l = l\pi$} 					& 
    \multirow{2}{*}{Sect.~\ref{sec_sub_EVP_outermu}} \\
    ($\mu=1$)		&   
    $(x/R)^{-1}$ 			&   
    ~~~ &   
			&   
			&   
			\\
			&   
    ~~~~if $x>R$			&   
     where $\nu=\bar{D}/(2\bar{m})$ &   
			&   
	$\beta=2$		&   
			\\
\hline   
    {outer $\mu$} 						& 
    $1$ if $x\le R$							& 
    \multirow{2}{*}{$\displaystyle\bar{n}\cot\bar{n}=\frac{1}{2}-\nu-\bar{m}\frac{K_{\nu-1}(\bar{m})}{K_{\nu}(\bar{m})}$}   			& 
    \multirow{3}{*}{$\displaystyle\frac{1}{2}$}					& 
    \multirow{2}{*}{$c_l=l\pi$} 					& 
    \multirow{2}{*}{Sect.~\ref{sec_sub_EVP_outermu}} \\
    ($\mu=2$)		&   
    $(x/R)^{-2}$ 			&   
    ~~~ &   
			&   
			&   
			\\
			&   
    ~~~~if $x>R$			&   
     where $\nu^2=1/4-\bar{D}$ &   
			&   
	$\beta=2$	&   
			\\
\hline   
    {inner $\mu$} 						& 
    $1-x/R$ 							& 
    \multirow{2}{*}{$\frac{\Bi(X_0) \Ai'(X_1)-\Ai(X_0) \Bi'(X_1)}{\Bi(X_0) \Ai(X_1)-\Ai(X_0) \Bi(X_1)} = -\frac{\bar{m}}{\bar{D}^{1/3}}$}   			& 
    \multirow{3}{*}{$\approx\displaystyle\frac{3\pi}{2}\left(l-\frac{5}{12}\right)$}					& 
    \multirow{2}{*}{$c_l = \displaystyle\frac{3(4l-1)\pi(1-\rhoe/\rhoi)}{8}$} 					& 
    \multirow{2}{*}{Sect.~\ref{sec_sub_EVP_innermu}} \\
    ($\mu=1$)		&   
    ~~~if $x\le R$ 			&   
     &   
			&   
			&   
			\\
			&   
     $0$ if $x>R$			&   
     where $X_0=-\bar{n}^2/\bar{D}^{2/3},  X_1=\bar{m}^2/\bar{D}^{2/3}$&   
			&   
	$\beta=2/3$		&   
			\\
\hline   
    {inner $\mu$} 						& 
    $1-x^2/R^2$ 							& 
    \multirow{2}{*}{$\displaystyle-\bar{m}=1-p+\frac{4p(\alpha+1/2)}{3}\frac{M(\alpha+3/2,5/2,p)}{M(\alpha+1/2,3/2,p)}$}   			& 
    \multirow{3}{*}{$\approx 4l-1$}					& 
    \multirow{2}{*}{$c_l = (4l-1)\sqrt{1-\rhoe/\rhoi}$} 					& 
    \multirow{2}{*}{ER88, also Sect.~\ref{sec_sub_EVP_innermu}} \\
    ($\mu=2$)		&   
    ~~~if $x\le R$ 			&   
     &   
			&   
			&   
			\\
			&   
     $0$ if $x>R$			&   
     where $p=\bar{D}^{1/2}, \alpha=1/4-\bar{n}^2/(4p)$&   
			&   
	$\beta=1$	&   
			\\
\hline   
    \multirow{3}{*}{exponential} 						& 
    \multirow{3}{*}{$\exp(-x/R)$} 							& 
    \multirow{3}{*}{$J_{2\bar{m}}(2\sqrt{\bar{D}})=0$}   			& 
    \multirow{3}{*}{$j_{0,l}/2$}					& 
    \multirow{2}{*}{$c_l=\displaystyle\frac{3(4l-1)\pi(1-\rhoe/\rhoi)}{8}$} 					& 
    \multirow{2}{*}{ER88 (C73, LG79), } \\
    		&   
     			&   
     &   
			&   
			&   
			\\
			&   
     			&   
       &   
			&   
	$\beta=2/3$		&   
	also Appendix~\ref{sec_app_EVP_exp}		\\
\hline   
    \multirow{2}{*}{Epstein} 						& 
    \multirow{2}{*}{${\rm sech}^2(x/R)$} 							& 
    \multirow{2}{*}{$\displaystyle\sqrt{\bar{D}+\frac{1}{4}}=\bar{m}+2l-\frac{1}{2}$}   			& 
    \multirow{2}{*}{$\sqrt{2l(2l-1)}$}					& 
    {$c_l = (4l-1)\sqrt{1-\rhoe/\rhoi}$} 					& 
    {ER88, CNW03, MR11 } \\
			&   
     			&   
       &   
			&   
	$\beta=1$	&   
	(LL65, LG79), also Appendix~\ref{sec_app_EVP_eps}		\\
\hline   
    \multirow{3}{*}{ } 						& 
    \multirow{3}{*}{$\displaystyle\frac{1}{(1+x/R)^2}$} 							& 
    \multirow{2}{*}{$K_\nu(\bar{m})=0$}   			& 
    \multirow{2}{*}{$\displaystyle\frac{1}{2}$}					& 
    \multirow{2}{*}{$\displaystyle c_l=\frac{3(4l-1)\pi(1-\rhoe/\rhoi)}{4}$} 					& 
    \multirow{2}{*}{LN15 (LG79)} \\
    		&   
     			&   
     &   
			&   
			&   
			\\
			&   
     			&   
     where $\nu^2 = 1/4-\bar{D}$ &   
			&   
	$\beta=2/3$	&   
	also Appendix~\ref{sec_app_EVP_invSqr}		\\	
\enddata
\tablenotetext{a}{\tiny Density profile: $\rho=\rhoe+(\rhoi-\rhoe)f(x)$. Definitions: $R$ -- half-width; $\vai$ and $\vae$ -- Alfv\'en speeds at $x=0$ and $\infty$;
   $\omega$ -- angular frequency; $k$ -- axial wavenumber; $\vph$ and $\vgr$ -- axial phase and group speeds; $l=1, 2, \cdots$ -- transverse order;
   $\bar{n}=kR\sqrt{\vph^2/\vai^2-1}>0$, $\bar{m}=kR\sqrt{1-\vph^2/\vae^2} \ge 0$, $\bar{D}=\bar{n}^2+\bar{m}^2 >0$.}
\tablenotetext{b}{\tiny Cutoff wavenumbers all take the form $\kcl R = d_l/\sqrt{\rhoi/\rhoe-1}$.}
\tablenotetext{c}{\tiny The phase speeds at large wavenumbers can all be described by $\vph^2/\vai^2 \approx 1+[c_l/(kR)]^{\beta}$, and consequently $\vgr^2/\vai^2 \approx 1+(1-\beta)[c_l/(kR)]^{\beta}$ .}
\tablenotetext{d}{\tiny References where the pertinent dispersion relation was given. Those in parentheses are for non-solar applications.
   C73:\citet{1973ApPhL..23..328C};  CNW03:\citet{2003A&A...409..325C}; ER82:\citet{1982SoPh...76..239E}; ER88:\citet{1988A&A...192..343E}; LG79:\citet{1979IJQE...15...14L}; LL65:\citet{1965qume.book.....L};
   LN15:\citet{2015ApJ...801...23L}; MR11:\citet{2011A&A...526A..75M}.
}
\tablenotetext{e}{\tiny When $kR \rightarrow 0$, $\vph^2/\vae^2 \approx 1-[(\rhoi/\rhoe-1)kR/(2l)]^2$.}
\end{deluxetable}

\clearpage

\clearpage

\clearpage

\clearpage

\clearpage

\clearpage

\clearpage

\clearpage

\clearpage

\clearpage

\clearpage

\end{document}